\documentclass[12pt,leqno]{article}

\title{Thermodynamics and collapse of self-gravitating \\
Brownian particles in $D$ dimensions}

\usepackage{amsthm,amsmath,amssymb,a4wide,psfig,epsf}
\def\mb#1{\setbox0=\hbox{$#1$}\kern-.025em\copy0\kern-\wd0
\kern-0.05em\copy0\kern-\wd0\kern-.025em\raise.0233em\box0}

\begin{document}

\author{Cl\'ement Sire and Pierre-Henri Chavanis}
\maketitle
\begin{center}
Laboratoire de Physique Quantique (UMR 5626 du CNRS), Universit\'e
Paul Sabatier,\\ 118, route de Narbonne, 31062 Toulouse, France\\
E-mail: {\it Clement.Sire{@}irsamc.ups-tlse.fr ~\&~
Chavanis{@}irsamc.ups-tlse.fr }

\vspace{0.5cm}
\end{center}

\begin{abstract}

We address the thermodynamics (equilibrium density profiles, phase
diagram, instability analysis...) and the collapse of a
self-gravitating gas of Brownian particles in $D$ dimensions, in both
canonical and microcanonical ensembles. In the canonical ensemble, we
derive the analytic form of the density scaling profile which decays
as $f(x)\sim x^{-\alpha}$, with $\alpha=2$. In the microcanonical
ensemble, we show that $f$ decays as $f(x)\sim x^{-\alpha_{\rm max}}$,
where $\alpha_{max}$ is a non-trivial exponent. We derive exact
expansions for $\alpha_{\rm max}$ and $f$ in the limit of large
$D$. Finally, we solve the problem in $D=2$, which displays rather
rich and peculiar features.

\end{abstract}

\section{Introduction}
\label{sec_introduction}

In an earlier paper \cite{charosi}, we studied a model of
self-gravitating Brownian particles confined within a
three-dimensional spherical box. We considered a high friction
limit in which the equations of the problem reduce to a
Smoluchowski-Poisson system with appropriate constraints ensuring
the conservation of energy (in the microcanonical ensemble) or
temperature (in the canonical ensemble) \cite{csr}. The
equilibrium states (maximum entropy states) correspond to
isothermal configurations which are known to exist only above a
critical energy or a critical temperature (see, e.g.,
\cite{pad}). When no hydrostatic equilibrium exists, we found
that the system generates a finite time singularity (i.e., the
central density becomes infinite in a finite time) and we derived
self-similar solutions describing the collapse. This study was
performed both in the microcanonical and canonical ensembles,
with emphasize on the inequivalence of ensembles for such a
nonextensive system. In the canonical ensemble, we showed that
the scaling exponent for the density is $\alpha=2$ and we
determined the invariant profile $f(x)$, satisfying $f(x)\sim
x^{-\alpha}$ for $x\rightarrow +\infty$, analytically. In the
microcanonical ensemble, the scaling exponent $\alpha\simeq
2.21...$ and the corresponding invariant profile $f(x)$ were
determined numerically.

In this paper, we propose to extend our analysis to a space of
arbitrary dimension $D$. The interest of this extension is
twofold. First, we shall consider an infinite dimension limit
$D\rightarrow +\infty$ in which the problem can be solved
analytically. In particular, it is possible to determine the
scaling exponent $\alpha(D)$ and the profile $f(x,D)$ in the
microcanonical ensemble by a systematic expansion procedure in
powers of $D^{-1}$ (Sec. \ref{sec_Micro}), while the canonical
value is always $\alpha=2$ and the profile can be calculated
exactly for any dimension (Sec. \ref{sec_Cane}). We show that,
already up to order $O(D^{-2})$, the results of the large $D$
expansion agree remarkably well with those found numerically for
$D=3$. Moreover, we show that the nature of the problem changes
at two particular dimensions $D=2$ and $D=10$. In Sec.
\ref{sec_statics}, we compute the equilibrium phase diagram as a
function of the dimension. For $2<D<10$, the $T-E$ curve has a
spiral shape like in $3D$. For $D>10$ and $D<2$, the $T-E$ curve
is monotonous. The dimension $D=2$ is {\it critical} and requires
a particular attention that is given in Sec. \ref{sec_twoD}. We
show that for $D=2$ the system generates a Dirac peak (containing
a finite fraction of mass) while for $D>2$, the central
singularity contains no mass (at the collapse time). The case
$D=2$ has interest in theoretical physics regarding 2D gravity
\cite{klb} and string theory \cite{polyakov} (in connection with
the Liouville field theory). It has also applications in the
physics of random surfaces \cite{cates} and random potentials
\cite{carpentier}, 2D turbulence \cite{onsager} and chemotaxis
\cite{murray} (for bacterial populations). Finally, the dynamical
equations considered in this paper and in Ref. \cite{csr} are
receiving a growing interest from mathematicians who established
rigorous results concerning the existence and unicity of
solutions for arbitrary domain shape without specific symmetry.
We refer to the papers of Rosier \cite{rosier} and Biler {\it et
al.} \cite{biler}, and references therein, for the connection of
our study with mathematical results.

\section{Equilibrium structure of isothermal spheres in dimension $D$}
\label{sec_statics}

\subsection{The maximum entropy principle}
\label{sec_maxent}

Consider a system of particles with mass $m$ interacting via Newtonian
gravity in a space of dimension $D$. The particles are enclosed within
a box of radius $R$ so as to prevent evaporation and make a
statistical approach rigorous. Let $f({\bf r},{\bf v},t)$ denote the
distribution function of the system, i.e. $f({\bf r},{\bf
v},t)d^{D}{\bf r}d^{D}{\bf v}$ gives the mass of particles whose
position and velocity are in the cell $({\bf r},{\bf v};{\bf
r}+d^{D}{\bf r},{\bf v}+d^{D}{\bf v})$ at time $t$. The integral of
$f$ over the velocity determines the spatial density
\begin{equation}
\rho=\int f \,d^{D}{\bf v}.
\label{maxent1}
\end{equation}
The total mass of the configuration is
\begin{equation}
M=\int \rho \,d^{D}{\bf r}.
\label{maxent2}
\end{equation}
In the meanfield approximation, the total energy of the system can be
expressed as
\begin{equation}
E={1\over 2}\int f v^{2}\,d^{D}{\bf r}d^{D}{\bf v}+{1\over
2}\int\rho\Phi \,d^{D}{\bf r}=K+W,
\label{maxent3}
\end{equation}
where $K$ is the kinetic energy and $W$ the potential energy. The
gravitational potential $\Phi$ is related to the density by the
Newton-Poisson equation
\begin{equation}
\Delta\Phi=S_{D}G\rho,
\label{maxent4}
\end{equation}
where $S_{D}$ is the surface of a unit sphere in a $D$-dimensional
space and $G$ is the constant of gravity. Finally, we introduce the
Boltzmann entropy
\begin{equation}
S=-\int f\ln f \,d^{D}{\bf r}d^{D}{\bf v},
\label{maxent5}
\end{equation}
and the free energy (more precisely the Massieu function)
\begin{equation}
J=S-\beta E,
\label{maxent6}
\end{equation}
where $\beta=1/T$ is the inverse temperature. If the system is
isolated, the equilibrium state maximizes the entropy $S$ at fixed
energy $E$ and mass $M$ ({microcanonical description}). Alternatively,
if the system is in contact with a heat bath which maintains its
temperature fixed, the equilibrium state maximizes the free energy $J$
at fixed mass $M$ and temperature $T$ ({canonical description}). It
can be shown that for systems interacting via a long-range potential
like gravity, this meanfield description is {\it exact} so that our
``thermodynamical approach'' is rigorous.

To solve this variational problem, we shall proceed in two steps. We
first maximize $S$ (resp. $J$) at fixed $M$, $E$ (resp. $T$) {\it and}
$\rho({\bf r})$. This yields the Maxwell distribution \begin{equation}
f={1\over (2\pi T)^{D/2}}\rho({\bf r})e^{-{v^{2}\over 2T}}.
\label{maxent7}
\end{equation}
It is now possible to express the energy and the entropy in terms of
$\rho({\bf r})$ as
\begin{equation}
E={D\over 2}MT+{1\over 2}\int \rho\Phi d^{D}{\bf r},
\label{maxent8}
\end{equation}
\begin{equation}
S={D\over 2}M\ln T-\int \rho\ln\rho d^{D}{\bf r},
\label{maxent9}
\end{equation}
where we have omitted unimportant constant terms in the entropy
(\ref{maxent9}). The entropy and the free energy are now functionals
of $\rho({\bf r})$ and we consider their maximization at fixed energy
or temperature. Introducing Lagrange multipliers to satisfy the
constraints, the critical points of $S$ or $J$ are given by the
Boltzmann distribution
\begin{equation}
\rho=Ae^{-\beta\Phi}.
\label{maxent10}
\end{equation}
Then, the equilibrium state is obtained by solving
the Boltzmann-Poisson equation
\begin{equation}
\Delta\Phi=S_{D}GAe^{-\beta\Phi},
\label{maxent11}
\end{equation}
and relating the Lagrange multipliers to the appropriate
constraints. Note that a similar variational problem occurs in the
context of two-dimensional turbulence ($D=2$) to characterize
large-scale vortices considered as maximum entropy structures
\cite{joyce,csr,csclass}.

It is easy to show that there is no global maximum of entropy at
fixed mass and energy for $D>2$ (see, e.g., Ref.~\cite{charosi}
for $D=3$). We can make the entropy diverge to $+\infty$ by
approaching an arbitrarily small fraction of particles in the
core ($M_{core}\ll M$) so that the potential energy goes to
$-\infty$. Since the total energy is conserved, the temperature
must rise to $+\infty$ and this leads to a divergence of the
entropy to $+\infty$. Note that if we collapse {\it all}
particles in the core, the entropy would diverge to $-\infty$.
Therefore, the formation of a Dirac peak is not thermodynamically
favorable in the microcanonical ensemble.  For $D=2$, there
exists a global entropy maximum for all energies. On the other
hand, there is no global maximum of free energy at fixed mass and
temperature for $D>2$ (see, e.g., Ref.~\cite{chavcano} for $D=3$)
and if $T<T_{c}=GM/4$ for $D=2$ (see Appendix \ref{sec_absence}).
We can make the free energy $J$ diverge to $+\infty$ by
collapsing all particles at $r=0$. Therefore, a canonical system
is expected to form a Dirac peak.  For $D=2$ and $T>T_{c}$, there
exists a global maximum of free energy. For $D<2$, there exists a
global maximum of entropy and free energy for all accessible
values of energy and temperature. We refer to
Refs.~\cite{kiessling,aly} for a rigorous proof of these results.
When no global maxima of entropy or free energy exist, we can
nevertheless look for local maxima since they correspond to
metastable states which can be relevant for the considered time
scales. Of course, the critical points of entropy at fixed $E$
and $M$ are the same as the critical point of free energy at
fixed $T$ and $M$. Only the onset of instability (regarding the
second order variations of $S$ or $J$ with appropriate
constraints) will differ from one ensemble to the other. For
$D=3$, this stability problem was considered by Antonov
\cite{antonov} and Padmanabhan \cite{pad2} in the microcanonical
ensemble and by Chavanis \cite{chavcano} in the canonical
ensemble, by solving an eigenvalue equation. It was also studied
by Lynden-Bell \& Wood \cite{lbw} and Katz \cite{katz78} by using
an extension of Poincar\'e theory of linear series of equilibria.
We shall give the generalization of these results in Sec.
\ref{sec_stab} to the case of a system of arbitrary dimension $D$.

\subsection{The $D$-dimensional Emden equation}
\label{sec_emden}

To determine the structure of isothermal spheres, we introduce the
function $\psi=\beta(\Phi-\Phi_{0})$, where $\Phi_{0}$ is the
gravitational potential at $r=0$. Then, the density field can be
written
\begin{equation}
\rho=\rho_{0}e^{-\psi},
\label{emden1}
\end{equation}
where $\rho_{0}$ is the central density. Introducing the notation
$\xi=(S_{D}G\beta\rho_{0})^{1/2}r$ and restricting ourselves to
spherically symmetrical configurations (which maximize the entropy for
a non-rotating system), the Boltzmann-Poisson Eq.~(\ref{maxent11})
reduces to the form
\begin{equation}
{1\over \xi^{D-1}}{d\over d\xi}\biggl (\xi^{D-1}{d\psi\over
d\xi}\biggr )=e^{-\psi},
\label{emden2}
\end{equation}
which is the $D$-dimensional generalization of the Emden equation
\cite{chandra}. For $D>2$, Eq.~(\ref{emden2}) has a simple explicit
solution, the singular sphere
\begin{equation}
e^{-\psi_{s}}={2(D-2)\over \xi^{2}}.
\label{emden3}
\end{equation}
The regular solution of Eq.~(\ref{emden2}) satisfying the boundary conditions
\begin{equation}
\psi=\psi'=0 \qquad {\rm at}\qquad \xi=0,
\label{emden4}
\end{equation}
must be computed numerically. For $\xi\rightarrow 0$, we can expand
the solution in Taylor series and we find that
\begin{equation}
\psi={1\over 2D}\xi^{2}-{1\over 8D(D+2)}\xi^{4}+
{1\over 24}{D+1\over D^{2}(D+2)(D+4)}\xi^{6}+...
\label{emden5}
\end{equation}
To obtain the asymptotic behavior of the solutions for $\xi\rightarrow
+\infty$, we note that the transformation $t=\ln\xi$, $\psi=2\ln\xi-z$
changes Eq.~(\ref{emden2}) in
\begin{equation}
{d^{2}z\over dt^{2}}+(D-2){dz\over dt}=-e^{z}+2(D-2).
\label{emden6}
\end{equation}
For $D>2$, this corresponds to the damped oscillations of a fictitious
particle in a potential $V(z)=e^{z}-2(D-2)z$, where $z$ plays the role
of the position and $t$ the role of time. For $t\rightarrow +\infty$,
the particle will come at rest at the bottom of the well at position
$z_{0}=\ln(2(D-2))$. Returning to original variables, we find that
\begin{equation}
e^{-\psi} \rightarrow {2(D-2)\over \xi^{2}}=e^{-\psi_{s}}, \qquad {\rm
for}\qquad \xi\rightarrow +\infty.
\label{emden7}
\end{equation}
Therefore, the regular solution of the Emden equation (\ref{emden2})
behaves like the singular solution for $\xi\rightarrow +\infty$. To
determine the next order correction, we set $z=z_{0}+z'$ with $z'\ll
1$. Keeping only terms that are linear in $z'$, Eq.~(\ref{emden6})
becomes
\begin{equation}
{d^{2}z'\over dt^{2}}+(D-2){dz'\over dt}+2(D-2)z'=0.
\label{emden8}
\end{equation}
The discriminant associated with this equation is
$\Delta=(D-2)(D-10)$. It exhibits two critical dimensions $D=2$ and
$D=10$. For $2<D<10$, we have
\begin{equation}
e^{-\psi}={2(D-2)\over \xi^{2}}\biggl\lbrace 1+{A\over \xi^{{D-2\over
2}}}\cos\biggl ({\sqrt{(D-2)(10-D)}\over 2}\ln\xi+\delta\biggr
)\biggr\rbrace, \qquad (\xi\rightarrow +\infty).
\label{emden9}
\end{equation}
The density profile (\ref{emden9}) intersects the singular solution
(\ref{emden3}) infinitely often at points that asymptotically increase
geometrically in the ratio $1:e^{2\pi/\sqrt{(D-2)(10-D)}}$ (see, e.g.,
Fig.~1 of Ref.~\cite{chavcano} for $D=3$). For $D \ge 10$, we have
\begin{equation}
e^{-\psi}={2(D-2)\over \xi^{2}}\biggl\lbrace 1+{1\over \xi^{{D-2\over
2}}}\biggl (A\xi^{\sqrt{(D-2)(10-D)}\over
2}+B\xi^{-\sqrt{(D-2)(10-D)}\over 2}\biggr ) \biggr\rbrace, \qquad
(\xi\rightarrow +\infty).
\label{emden10}
\end{equation}
For $D=2$, Eq.~(\ref{emden6}) can be solved explicitly and we get
\begin{equation}
e^{-\psi}={1\over (1+{1\over 8}\xi^{2})^{2}}.
\label{emden11}
\end{equation}
This result has been found by various authors in different fields
(see, e.g., \cite{klb,caglioti}).  Note that $e^{-\psi}\sim \xi^{-4}$
at large distances instead of the usual $\xi^{-2}$ behavior obtained
for $D>2$. This implies that the mass of an unbounded isothermal
sphere is finite in $D=2$, although it is infinite for $D>2$.

For $D<2$, we can neglect the r.h.s. in Eq.~(\ref{emden7}) and we get
\begin{equation}
e^{-\psi}\sim e^{-A_{D}\xi^{2-D}}, \qquad (\xi\rightarrow +\infty),
\label{emden12}
\end{equation}
where $A_{D}$ is a constant depending on the dimension $D$. For $D=1$,
Eq.~(\ref{emden2}) can be solved exactly, yielding the result
\begin{equation}
e^{-\psi}={1\over \cosh^{2}(\xi/\sqrt{2})},
\label{emden13}
\end{equation}
establishing $A_{1}=\sqrt{2}$.

\subsection{The Milne variables}
\label{sec_milne}

\begin{figure}
\centerline{
\psfig{figure=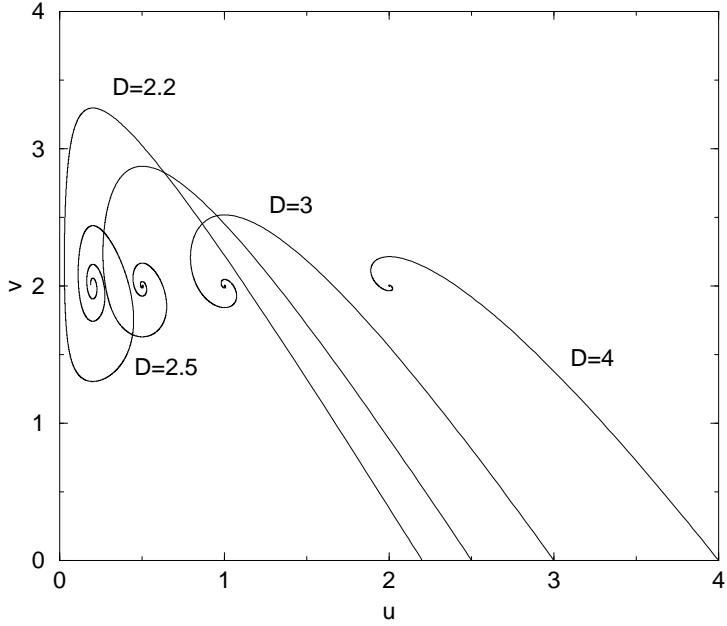,angle=0,height=8.5cm}}
\caption{The solutions of the Emden equation in the $(u,v)$ plane
for systems with dimension $2<D<10$.}
\label{uv3}
\end{figure}

\begin{figure}
\centerline{
\psfig{figure=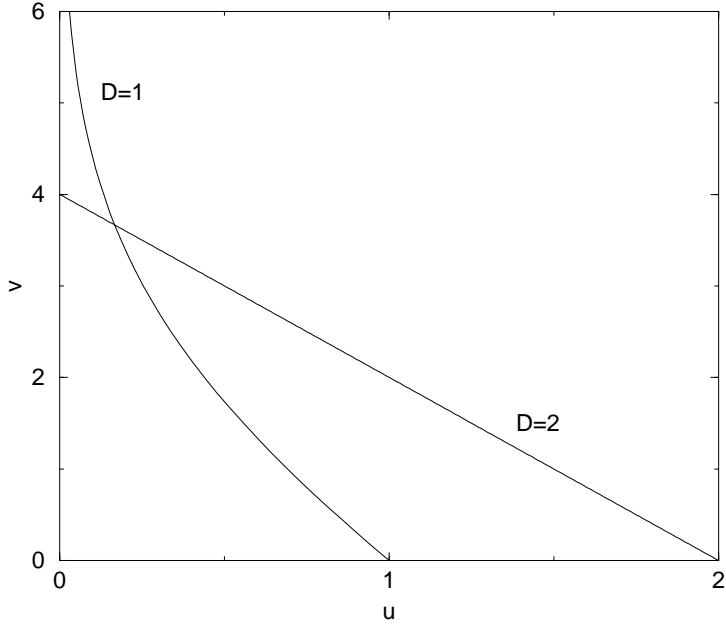,angle=0,height=8.5cm}}
\caption{The solutions of the Emden equation in the $(u,v)$ plane for
systems with dimension $D=1$ and $D=2$.}
\label{uvD1et2}
\end{figure}

As is well-known \cite{chandra}, isothermal spheres satisfy a homology
theorem: if $\psi(\xi)$ is a solution of the Emden equation, then
$\psi(A\xi)-2\ln A$ is also a solution, with $A$ an arbitrary
constant. This means that the profile of isothermal configurations is
always the same (characterized intrinsically by the function $\psi$),
provided that the central density and the typical radius are rescaled
appropriately. Because of this homology theorem, the second order
differential equation (\ref{emden2}) can be reduced to a {\it first
order} differential equation for the Milne variables
\begin{equation}
u={\xi e^{-\psi}\over \psi'}\qquad {\rm and}\qquad v=\xi\psi'.
\label{milne1}
\end{equation}
Taking the logarithmic derivative of $u$ and $v$ with respect to $\xi$
and using Eq.~(\ref{emden2}), we get
\begin{equation}
{1\over u}{du\over d\xi}={1\over\xi}(D-v-u),
\label{milne2}
\end{equation}
\begin{equation}
{1\over v}{dv\over d\xi}={1\over\xi}(2-D+u).
\label{milne3}
\end{equation}
Taking the ratio of the foregoing equations, we obtain
\begin{equation}
{u\over v}{dv\over du}={2-D+u\over D-u-v}.
\label{milne3b}
\end{equation}
The solution curve in the $(u,v)$ plane is plotted in Fig.~\ref{uv3}
for different values of $D$. The curve is parameterized by $\xi$. It
starts from the point $(u,v)=(D,0)$ with a slope
$(dv/du)_{0}=-(D+2)/D$ corresponding to $\xi=0$. The points of
horizontal tangent are determined by $u=D-2$ and the points of
vertical tangent by $u+v=D$. These two lines intersect at
$(u_{s},v_{s})=(D-2,2)$, which corresponds to the singular solution
(\ref{emden3}). For $2<D<10$, the solution curve spirals indefinitely
around the point $(u_{s},v_{s})$.  For $D\ge 10$, the curve reaches
the point $(u_{s},v_{s})$ without spiraling. For $D=2$, we have the
explicit solution $v=2(2-u)$ so that $(u,v)\rightarrow (0,4)$ for
$\xi\rightarrow +\infty$. For $D<2$, $(u,v)\rightarrow (0,+\infty)$
for $\xi\rightarrow +\infty$ (see Fig.~\ref{uvD1et2}). More precisely,
\begin{equation}
{u e^{v\over 2-D}\over v^{D\over 2-D}}\sim \omega_{D}, \qquad
(\xi\rightarrow +\infty),
\label{milne4}
\end{equation}
where we have defined $\omega_{D}=1/(A_{D}(2-D)^{2\over 2-D})$. For
$D=1$, $\omega_{1}=1/\sqrt{2}$.

\subsection{The thermodynamical parameters}
\label{sec_thermo}

For bounded isothermal systems, the solution of Eq.~(\ref{emden2}) is
terminated by the box at a normalized radius given by
$\alpha=(S_{D}\beta\rho_{0})^{1/2}R$. We shall now relate the
parameter $\alpha$ to the temperature and energy. According to the
Poisson equation (\ref{maxent4}), we have
\begin{equation}
GM=\int G\rho d^{D}{\bf r}=\int_{0}^{R}S_{D}G\rho
r^{D-1}dr=\int_{0}^{R}{d\over dr}\biggl (r^{D-1}{d\Phi\over dr}\biggr
)dr=\biggl (r^{D-1}{d\Phi\over dr}\biggr )_{r=R}.
\label{thermo1}
\end{equation}
Introducing the dimensionless variables defined previously (using
$r/R=\xi/\alpha$), we get
\begin{equation}
\eta\equiv {\beta GM\over R^{D-2}}=\alpha\psi'(\alpha).
\label{thermo2}
\end{equation}
We note that, for $D=2$, the parameter $\eta$ is independent on
$R$. This is a consequence of the logarithmic form of the Newtonian
potential in two dimensions.

The computation of the energy is a little more intricate. First, using
an extension of the potential tensor theory developed by Chandrasekhar
(see, e.g., Ref.~\cite{bt}), we can show that the potential energy in
$D$-dimensions can be written
\begin{equation}
W=-{1\over D-2}\int\rho\ {\bf r}\cdot\nabla\Phi d^{D}{\bf r},
\label{thermo3}
\end{equation}
for $D\neq 2$. Now, the Boltzmann-Poisson equation (\ref{maxent11}) is
equivalent to the condition of hydrostatic equilibrium
\begin{equation}
\nabla p=-\rho\nabla\Phi,
\label{thermo4}
\end{equation}
with an equation of state $p=\rho T$.  Substituting this relation in
Eq.~(\ref{thermo3}) and integrating by parts, we obtain
\begin{equation}
2K+(D-2)W=DV_{D}R^{D}p(R),
\label{thermo5}
\end{equation}
where $V_{D}=S_{D}/D$ is the volume of a hypersphere with unit
radius. Eq.~(\ref{thermo5}) is the form of the Virial theorem in
$D$-dimension. The total energy $E=K+W$ can be written
\begin{equation}
E={D-4\over D-2}K+{D\over D-2}V_{D}R^{D}p(R).
\label{thermo6}
\end{equation}
Expressing the pressure in terms of the Emden function, using $p=\rho
T$ and Eq.~(\ref{emden1}), and using Eq.~(\ref{thermo2}) to eliminate
the temperature, we finally obtain
\begin{equation}
\Lambda\equiv -{ER^{D-2}\over GM^{2}}={D(4-D)\over 2(D-2)}
{1\over\alpha\psi'(\alpha)}-{1\over
D-2}{e^{-\psi(\alpha)}\over\psi'(\alpha)^{2}}.
\label{thermo7}
\end{equation}

It turns out that the normalized temperature and the normalized energy
can be expressed very simply in terms of the values of the Milne
variables at the normalized box radius. Indeed, writing
$u_{0}=u(\alpha)$ and $v_{0}=v(\alpha)$ and using
Eqs.~(\ref{thermo2})(\ref{thermo7}), we get
\begin{equation}
\eta=v_{0},
\label{thermo8}
\end{equation}
\begin{equation}
\Lambda={1\over v_{0}}\biggl\lbrack {D(4-D)\over 2(D-2)}
-{u_{0}\over D-2}\biggr\rbrack.
\label{thermo9}
\end{equation}

\begin{figure}
\centerline{
\psfig{figure=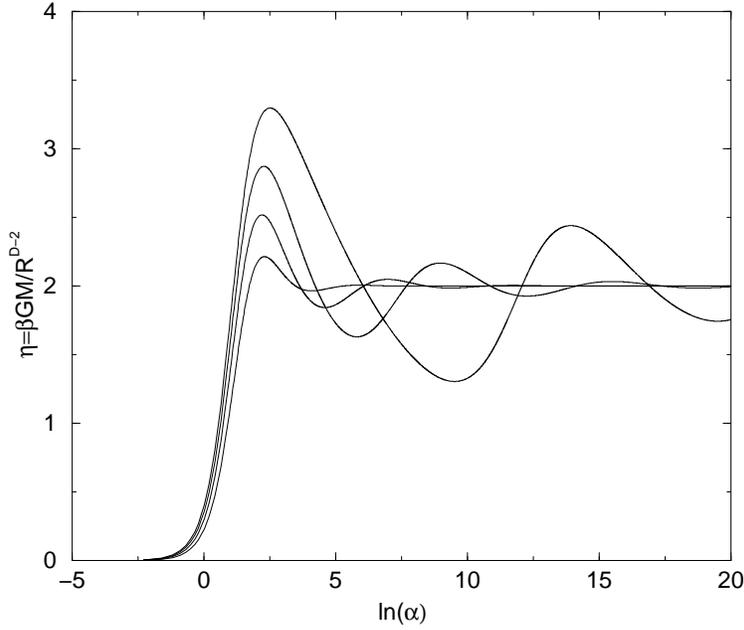,angle=0,height=8.5cm}}
\caption{Evolution of the inverse temperature $\eta$ along the series
of equilibrium (parameterized by $\alpha$) for $2<D<10$. The curves
correspond to $D=4,3,2.5,2.2$ from bottom to top.}
\label{alphaetaall}
\end{figure}

\begin{figure}
\centerline{
\psfig{figure=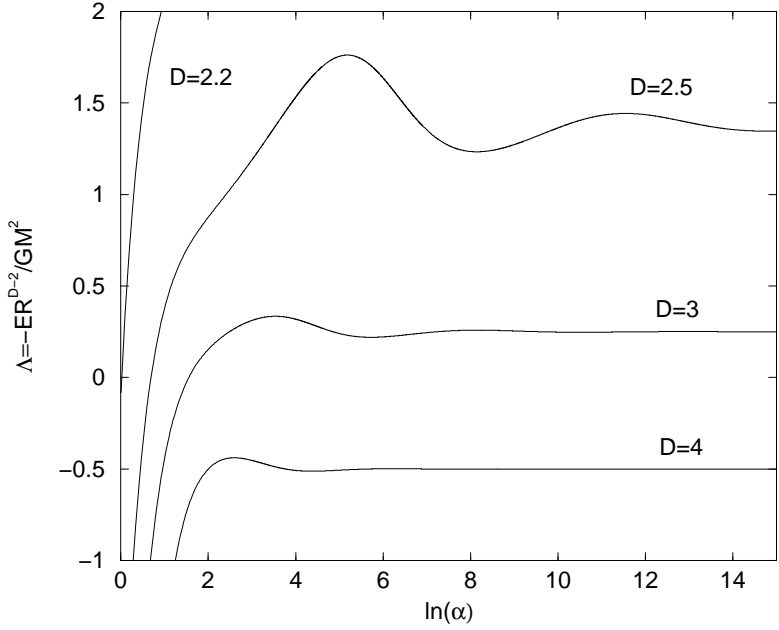,angle=0,height=8.5cm}}
\caption{Evolution of the energy $\Lambda$ along the series of
equilibria (parameterized by $\alpha$) for $2<D<10$. }
\label{alphaLambdaall}
\end{figure}

\begin{figure}
\centerline{
\psfig{figure=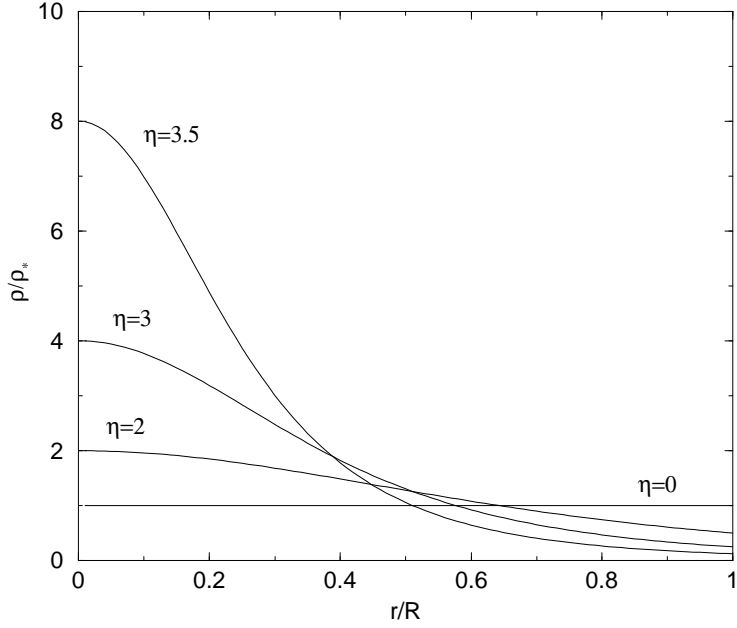,angle=0,height=8.5cm}}
\caption{Equilibrium density profile of a two-dimensional self-gravitating
system as a function of the inverse temperature $\eta$. For $\eta=0$,
the density is uniform. For $\eta\rightarrow \eta_{c}=4$, the density
tends to a Dirac peak. For $\eta>\eta_{c}$, there is no equilibrium
state.}
\label{profil2D}
\end{figure}

The curves $\eta(\alpha)$, $\Lambda(\alpha)$ are plotted in
Figs.~\ref{alphaetaall}-\ref{alphaLambdaall}. For $2<D<10$, they
exhibit damped oscillations toward the values $\eta_{s}=2$ and
$\Lambda_{s}=1/(D-2)-D/4$, corresponding to the singular solution
(\ref{emden3}). For $D\ge 10$ the curves are monotonous. For $D=2$, we
have explicitly
\begin{equation}
\eta={\alpha^{2}\over 2(1+{1\over 8}\alpha^{2})}, \qquad
\Lambda={2\over\alpha^{2}}\biggl (1+{\alpha^{2}\over 8}\biggr )\biggl\lbrace
{8\over\alpha^{2}}\biggl (1+{\alpha^{2}\over 8}\biggr )\ln\biggl
(1+{\alpha^{2}\over 8}\biggr )-2\biggr\rbrace.
\label{thermo10}
\end{equation}
The expression of the energy has been obtained directly from
Eq.~(\ref{maxent8}) with the boundary condition $\Phi(R)=0$. The
inverse temperature increases monotonically with $\alpha$ up to the
value $\eta_{c}=4$. Using Eq.~(\ref{emden11}) and returning to the
original variables, we can write the density profile in the form
\begin{equation}
\rho={4M\over \pi R^{2}(4-\eta)(1+{\eta\over 4-\eta}{r^{2}\over R^{2}})^{2}}.
\label{thermo11}
\end{equation}
This density profile is represented in Fig.~\ref{profil2D} for
different temperatures. At the critical inverse temperature
$\eta_{c}=4$, all the particles are concentrated at the center of the
domain. The density profile approaches the Dirac distribution
\begin{equation}
\rho({\bf r})\rightarrow M\delta({\bf r})\qquad {\rm for}\qquad
\eta\rightarrow \eta_{c}=4,
\label{thermo12}
\end{equation}
which has an infinite (negative) energy.

For $D<2$, the curves $\eta(\alpha)$ and $\Lambda(\alpha)$ tend to
$+\infty$ and $0$ as $\alpha\rightarrow +\infty$. For $D=1$, we have
explicitly
\begin{equation}
\eta=2\sqrt{2}\alpha\tanh(\alpha/\sqrt{2}),\qquad
\Lambda=-{3\over 4\sqrt{2}}{1\over\alpha\tanh(\alpha/\sqrt{2})}
+{1\over 8}{1\over \sinh^{2}(\alpha/\sqrt{2})}.
\label{thermo13}
\end{equation}
Note that according to Eq.~(\ref{thermo3}), the
energy is necessarily positive for $D<2$, so the region $\Lambda\ge 0$
is forbidden.  Returning to original variables, the density profile is
given by
\begin{equation}
\rho={M\over 4\sqrt{2}R}\ {\alpha\over\tanh(\alpha/\sqrt{2})}\
{1\over\cosh^{2}\bigl ({\alpha r\over \sqrt{2}R}\bigr)},
\label{thermo14}
\end{equation}
where we recall that $S_{1}=2$. For $\alpha\rightarrow +\infty$, the
profile tends to a Dirac peak $M\delta(r)$.

\begin{figure}
\centerline{
\psfig{figure=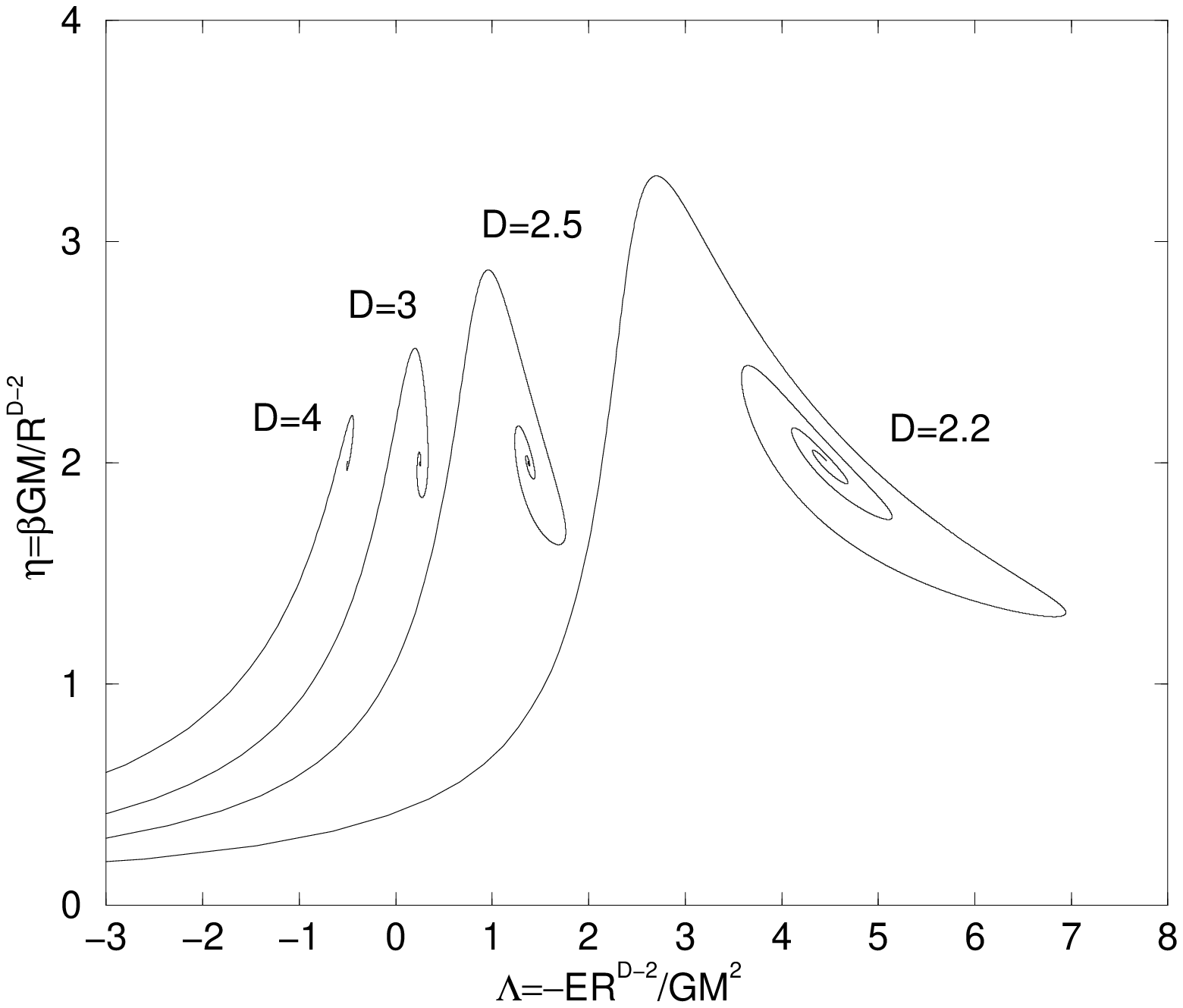,angle=0,height=8.5cm}}
\caption{Equilibrium phase diagram giving the inverse temperature
$\eta$ as a function of minus the energy $\Lambda$ for systems with
dimension $2<D<10$.}
\label{Lambdaeta3}
\end{figure}

\begin{figure}
\centerline{
\psfig{figure=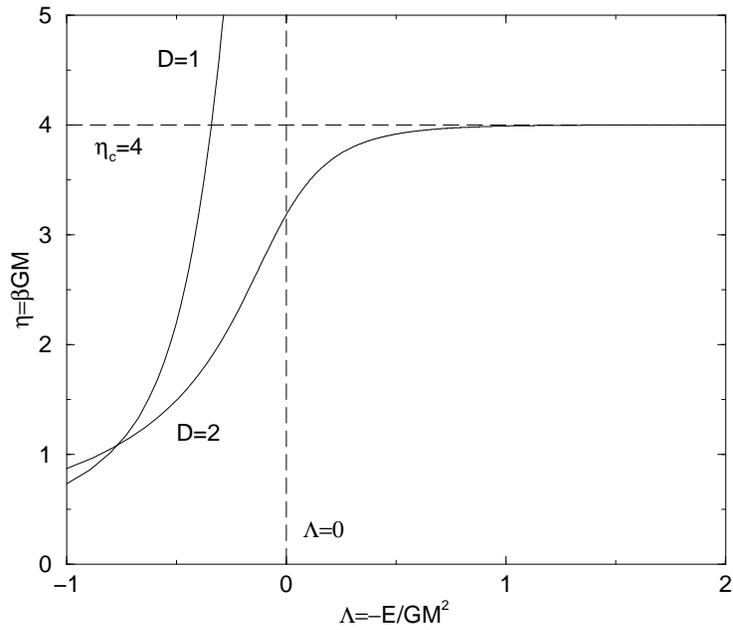,angle=0,height=8.5cm}}
\caption{Equilibrium phase diagram for two-dimensional self-gravitating
systems. For infinitely negative energies, the inverse temperature
tends to the value $\eta_{c}=4$.}
\label{Lambdaeta2}
\end{figure}

In Figs.~\ref{Lambdaeta3}-\ref{Lambdaeta2}, we have plotted the
equilibrium phase diagram $\Lambda-\eta$, giving the temperature as a
function of the energy, for different dimensions $D$. For $2<D<10$,
the curve spirals around the limit point $(\Lambda_{s},\eta_{s})$
corresponding to the singular solution. For $D\ge 10$, the curve is
monotonous until the limit point. For $D=2$, the curve is explicitly
given by
\begin{equation}
\Lambda={1\over \eta}\biggl\lbrack {4\over\eta}\ln\biggl ({4\over 4-\eta}
\biggr )-2\biggr\rbrack,
\label{thermo12a}
\end{equation}
and is represented in Fig.~\ref{Lambdaeta2}, together with the case
$D=1$.

We stress that the preceding results, obtained in the meanfield
approximation, are exact in the thermodynamical limit $N\rightarrow
+\infty$ such that $\eta$ and $\Lambda$ are kept fixed. If the box
radius is given, this implies that $T\sim N$ and $E\sim
N^{2}$. Alternatively, if the temperature and the energy per particle
are given, the thermodynamical limit is such that $N\rightarrow
+\infty$ with $N/R^{D-2}$ constant (for $D>2$).

\subsection{The minimum temperature and minimum energy}
\label{sec_min}

\begin{figure}
\centerline{
\psfig{figure=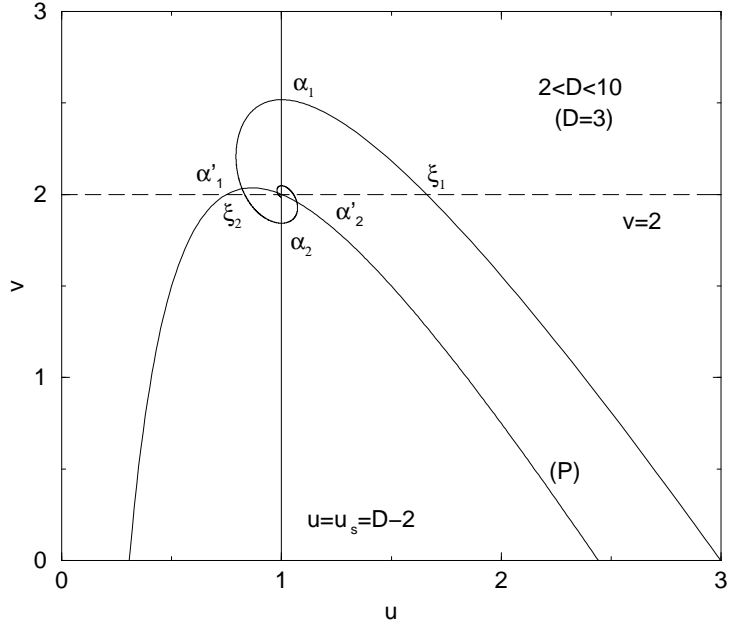,angle=0,height=8.5cm}}
\caption{Location of the turning points of energy and temperature in
the $(u,v)$ plane for systems with dimension $2<D<10$. The
construction is made explicitly for $D=3$, which corresponds to the
case extensively studied in Refs. \cite{pad2,chavcano}. The dashed
line $v=2$ determines the location of the nodes of the density
profiles that trigger the instabilities (see Sec. \ref{sec_stab}).}
\label{uvparabole3}
\end{figure}

For $2<D<10$, the curve $\eta(\alpha)$ presents an extremum at points
$\alpha_{n}$ such that $d\eta/d\alpha(\alpha_{n})=0$. Using
Eqs.~(\ref{thermo8}) and (\ref{milne3}), we find that this condition
is equivalent to
\begin{equation}
u_{0}=D-2=u_{s}.
\label{min1}
\end{equation}
Since the curve $u=u_{s}$ passes through the center of the spiral in
the $(u,v)$ plane, this determines an infinity of solutions (see
Fig.~\ref{uvparabole3}), one at each extremum of $v$ (since
$\eta=v_{0}$). Asymptotically, the $\alpha_{n}$ follow a geometric
progression (see Ref.~\cite{chavcano} for more details):
\begin{equation}
\alpha_{n}\sim e^{2\pi n/\sqrt{(D-2)(10-D)}}, \qquad (n\rightarrow
+\infty,\ {\rm integer}).
\label{min2}
\end{equation}
In Fig.~\ref{alphaetaall}, we see that an equilibrium state exists
only for
\begin{equation}
\eta={\beta GM\over R^{D-2}}\le \eta(\alpha_{1}),\qquad (2<D<10).
\label{min3}
\end{equation}
This determines a maximum mass (for given $T$ and $R$) or a minimum
temperature (for given $M$ and $R$) beyond which no equilibrium state
is possible. In that case, the system is expected to undergo an {\it
isothermal collapse} (see Sect. \ref{sec_Cane}). For $D=2$ and for
$D\ge 10$, the $\eta(\alpha)$ curve is monotonous. An equilibrium
state exists only provided that
\begin{equation}
\eta= {\beta GM}\le \eta_{c}=4,\qquad (D=2),
\label{min4}
\end{equation}
\begin{equation}
\eta={\beta GM\over R^{D-2}}\le \eta_{s}=2,\qquad (D\ge 10).
\label{min4b}
\end{equation}

We get comparable results for the energy. For $2<D<10$, the curve
$\Lambda(\alpha)$ presents an extremum at points $\alpha'_{n}$ such
that $d\Lambda/d\alpha(\alpha'_{n})=0$. Using Eqs.~(\ref{thermo9}) and
(\ref{milne2})(\ref{milne3}), we find that this condition is
equivalent to
\begin{equation}
4u_{0}^{2}+2u_{0}v_{0}+(D^{2}-8D+4)u_{0}+D(D-2)(4-D)=0.
\label{min5}
\end{equation}
We can check that the limit point $(u_{s},v_{s})$ is solution of this
equation.  Therefore, the intersection of the parabola $(P)$ defined
by Eq. (\ref{min5}) with the spiral in the $(u,v)$ plane determines an
infinity of points $\alpha'_{n}$ at which the energy is extremum (see
Fig.~\ref{uvparabole3}).  On Fig.~\ref{alphaLambdaall}, we see that an
equilibrium state exists only for
\begin{equation}
\Lambda={-ER^{D-2}\over GM^{2}}\le \Lambda(\alpha'_{1}),\qquad (2<D<10).
\label{min6}
\end{equation}
This determines a minimum energy (for given $M$ and $R$) or a maximum
radius (for given $M$ and $E$) beyond which no equilibrium state
exists. In that case, the system is expected to collapse and overheat;
this is called {\it gravothermal catastrophe} (see
Sect. \ref{sec_Micro}). For $D\ge 10$, the curve $\Lambda(\alpha)$ is
monotonous. An equilibrium state exist only for
\begin{equation}
\Lambda={-ER^{D-2}\over GM^{2}}\le \Lambda_{s}={1\over D-2}-{D\over 4},
\qquad (D\ge 10).
\label{min7}
\end{equation}
For $D=2$, there exists an equilibrium state for each value of energy
(see Fig.~\ref{Lambdaeta2}): there is no gravothermal catastrophe in
the microcanonical ensemble in two-dimensions \cite{klb}. For $D<2$,
there exists an equilibrium state for all accessible values of energy
and temperature.

\subsection{The thermodynamical stability}
\label{sec_stab}

We now study the thermodynamical stability of self-gravitating systems
in various dimensions. We start by the canonical ensemble which is
simpler in a first approach. A critical point of free energy at fixed
mass and temperature is a local maximum if, and only if, the second
order variations
\begin{equation}
\delta^{2}J=-\int {(\delta\rho)^{2}\over 2\rho}d^{D}{\bf r}-{1\over 2T}\int
\delta\rho\delta\Phi d^{D}{\bf r},
\label{stab1}
\end{equation}
are negative for any variation $\delta\rho$ that conserves mass to
first order, i.e.
\begin{equation}
\int\delta\rho d^{D}{\bf r}=0.
\label{stab2}
\end{equation}
This is the condition of thermodynamical stability in the canonical
ensemble. Introducing the function $q(r)$ by the relation
\begin{equation}
\delta\rho={1\over S_{D}r^{D-1}}{dq\over dr},
\label{stab3}
\end{equation}
and following a procedure similar to the one adopted in
Ref.~\cite{chavcano}, we can put the second order variations of free
energy in the quadratic form
\begin{equation}
\delta^{2}J={1\over 2}\int_{0}^{R}dr q\biggl\lbrack {G\over Tr^{D-1}}
+{d\over dr}\biggl ({1\over S_{D}\rho r^{D-1}}{d\over dr}\biggr
)\biggr \rbrack q.
\label{stab4}
\end{equation}
The second order variations of free energy can be positive (implying
instability) only if the differential operator which occurs in the
integral has positive eigenvalues. We need therefore to consider the
eigenvalue problem
\begin{equation}
\biggl\lbrack {d\over dr}\biggl ({1\over S_{D}\rho r^{D-1}}
{d\over dr}\biggr )+{G\over Tr^{D-1}}\biggr \rbrack
q_{\lambda}(r)=\lambda q_{\lambda}(r).
\label{stab5}
\end{equation}
with $q_{\lambda}(0)=q_{\lambda}(R)=0$. If all the eigenvalues
$\lambda$ are negative, then the critical point is a {maximum} of free
energy. If at least one eigenvalue is positive, the critical point is
an unstable saddle point. The point of marginal stability, i.e. the
value of $\alpha$ in the series of equilibria $\eta(\alpha)$ at which
the solutions pass from local maxima of free energy to unstable saddle
points, is determined by the condition that the largest eigenvalue is
equal to zero ($\lambda=0$). We thus have to solve the differential
equation
\begin{equation}
{d\over dr}\biggl ({1\over S_{D}\rho r^{D-1}}{dF\over dr}\biggr
)+{GF\over Tr^{D-1}}=0,
\label{stab6}
\end{equation}
with $F(0)=F(R)=0$. Introducing the dimensionless variables defined
previously, we can rewrite this equation in the form
\begin{equation}
{d\over d\xi}\biggl ({e^{\psi}\over \xi^{D-1}}{dF\over d\xi}\biggr
)+{F(\xi)\over \xi^{D-1}}=0,
\label{stab7}
\end{equation}
with $F(0)=F(\alpha)=0$. If
\begin{equation}
{\cal L}\equiv {d\over d\xi}\biggl ({e^{\psi}\over \xi^{D-1}}{d\over
d\xi}\biggr )+{1\over \xi^{D-1}}
\label{stab7a}
\end{equation}
denotes the differential operator which occurs in Eq.~(\ref{stab7}),
we can check by using the Emden Eq.~(\ref{emden2}) that
\begin{equation}
{\cal L}(\xi^{D-1}\psi')=\psi',\qquad {\cal L}(\xi^{D}e^{-\psi})=(D-2)\psi'.
\label{stab7b}
\end{equation}
Therefore, the general solution of Eq.~(\ref{stab7}) satisfying the
boundary conditions at $\xi=0$ is
\begin{equation}
F(\xi)=c_{1}(\xi^{D}e^{-\psi}-(D-2)\xi^{D-1}\psi').
\label{stab8}
\end{equation}
Using Eq.~(\ref{stab8}) and
introducing the Milne variables (\ref{milne1}), the condition
$F(\alpha)=0$ can be written
\begin{equation}
u_{0}=D-2.
\label{stab8a}
\end{equation}
This relation determines the points at which a new eigenvalue becomes
positive (crossing the line $\lambda=0$). Comparing with
Eq.~(\ref{min1}), we see that a mode of stability is lost each time
that $\eta$ is extremum in the series of equilibria, in agreement with
the turning point criterion of Katz \cite{katz78} in the canonical
ensemble. In particular, the series becomes unstable at the point of
minimum temperature (or maximum mass) $\alpha_{1}$. Secondary modes of
instability appear at values $\alpha_{2}$, $\alpha_{3}$,... We obtain
the same results by considering the dynamical stability of isothermal
gaseous spheres with respect to the Navier-Stokes equations (see
Ref.~\cite{chavcano} for $D=3$). Therefore, dynamical and
thermodynamical stability criteria coincide for isothermal gaseous
spheres.

According to Eq.~(\ref{stab3}), the perturbation profile that triggers
a mode of instability at the critical point $\lambda=0$ is given by
\begin{equation}
{\delta\rho\over\rho_{0}}={1\over S_{D}\xi^{D-1}}{dF\over d\xi},
\label{stab9}
\end{equation}
where $F(\xi)$ is given by Eq.~(\ref{stab8}). Introducing the Milne
variables (\ref{milne1}), we get
\begin{equation}
{\delta\rho\over\rho}={c_{1}\over S_{D}}(2-v).
\label{stab9b}
\end{equation}
The density perturbation $\delta\rho$ becomes zero at point(s)
$\xi_{i}$ such that $v(\xi_{i})=2$. The number of zeros is therefore
given by the number of intersections between the spiral in the $(u,v)$
plane and the line $v=2$ (see Fig.~\ref{uvparabole3}). For the $n$-th
mode of instability we need to follow the spiral to the $n$-th
extremum of $v$ (since $\alpha_{n}$ corresponds to an extremum of
$\eta$, hence $v$). Therefore, the density perturbation $\delta\rho$
corresponding to the $n$-th mode of instability has $n$ zeros
$\xi_{1}, \xi_{2},..., \xi_{n}<\alpha_{n}$. Asymptotically, the zeros
follow a geometric progression with ratio
$e^{2\pi/\sqrt{(D-2)(10-D)}}$ \cite{chavcano}.

In the microcanonical ensemble, the condition of thermodynamical
stability requires that the equilibrium state is an entropy {maximum}
at fixed mass and energy. This condition can be written
\begin{equation}
\delta^{2}S=-\int {(\delta\rho)^{2}\over 2\rho}d^{D}{\bf r}-{1\over 2T}
\int \delta\rho\delta\Phi d^{D}{\bf r}-{1\over DMT^{2}}\biggl
(\int\Phi\delta\rho d^{D}{\bf r}\biggr )^{2}<0,
\label{stab10}
\end{equation}
for any variation $\delta\rho$ that conserves mass to first order (the
conservation of energy has already been taken into account in
obtaining Eq.~(\ref{stab10})). Now, following a procedure similar to
that of Ref.~\cite{pad} in $D=3$, the second variations of entropy can
be put in a quadratic form
\begin{equation}
\delta^{2}S=\int_{0}^{R}\int_{0}^{R}dr dr' q(r)K(r,r')q(r'),
\label{stab11}
\end{equation}
with
\begin{equation}
K(r,r')=-{1\over DMT^{2}}{d\Phi\over dr}(r){d\Phi\over dr}(r')+{1\over
2}\delta(r-r')\biggl \lbrack {G\over Tr^{D-1}}+{d\over dr}\biggl
({1\over S_{D}\rho r^{D-1}}{d\over dr}\biggr )\biggr\rbrack.
\label{stab12}
\end{equation}
The problem of stability can therefore be reduced to the study of the
eigenvalue equation
\begin{equation}
\int_{0}^{R}dr' K(r,r')F_{\lambda}(r')=\lambda F_{\lambda}(r),
\label{stab13}
\end{equation}
with $F_{\lambda}(0)=F_{\lambda}(R)=0$. The point of marginal
stability ($\lambda=0$) will be determined by solving the differential
equation
\begin{equation}
{d\over dr}\biggl ({1\over S_{D}\rho r^{D-1}}{dF\over dr}\biggr
)+{GF\over Tr^{D-1}}={2V\over DMT^{2}}{d\Phi\over dr}(r),
\label{stab14}
\end{equation}
with
\begin{equation}
V=\int_{0}^{R}{d\Phi\over dr}(r')F(r')dr'.
\label{stab15}
\end{equation}
Introducing the dimensionless variables defined previously, this is
equivalent to
\begin{equation}
{d\over d\xi}\biggl ({e^{\psi}\over \xi^{D-1}}{dF\over d\xi}\biggr
)+{F\over \xi^{D-1}}=\chi{d\psi\over d\xi},
\label{stab16}
\end{equation}
with
\begin{equation}
\chi={2\over D\alpha^{D-1}\psi'(\alpha)}\int_{0}^{\alpha}{d\psi\over d\xi}
(\xi')F(\xi')d\xi',
\label{stab17}
\end{equation}
and $F(0)=F(\alpha)=0$. Using the identities (\ref{stab7b}), we can
check that the general solution of Eq.~(\ref{stab16}) satisfying the
boundary conditions for $\xi=0$ and $\xi=\alpha$ is
\begin{equation}
F(\xi)={\chi\over D-2-u_{0}}\biggl
(\xi^{D}e^{-\psi}-(D-2)\xi^{D-1}\psi'\biggr )+{\chi}\xi^{D-1}\psi',
\label{stab18}
\end{equation}
The point of marginal stability is then obtained by substituting the
solution (\ref{stab18}) in Eq.~(\ref{stab17}). Using the identities
\begin{equation}
\int_{0}^{\alpha}\psi'\xi^{D}e^{-\psi}d\xi=\alpha^{D-1}\psi'(\alpha)
(D-{u_{0}} ),
\label{stab19}
\end{equation}
\begin{equation}
(D-2)\int_{0}^{\alpha}\xi^{D-1}(\psi')^{2}d\xi=\alpha^{D-1}\psi'(\alpha)
(2D-2u_{0}-v_{0}),
\label{stab20}
\end{equation}
which result from simple integrations by parts and from the properties
of the Emden equation (\ref{emden2}), it is found that the point of
marginal stability is determined by the condition
(\ref{min5}). Therefore, the series of equilibria becomes unstable at
the point of minimum energy in agreement with the turning point
criterion of Katz \cite{katz78} in the microcanonical ensemble. The
structure of the perturbation that triggers the instability can be
determined with the graphical construction described in
Ref.~\cite{pad2}. It is found that the first mode of instability has a
core-halo structure (i.e., two nodes) in the microcanonical ensemble,
unlike the first mode of instability in the canonical ensemble \cite{chavcano}.

The thermodynamical stability analysis presented in this section also
shows that the equilibrium states for $D\le 2$ and $D\ge 10$ are
always stable since the series of equilibria do not present turning
points of energy or temperature.

\section{Dynamics of self-gravitating Brownian particles in dimension $D$}
\label{sec_dynamics}

\subsection{The Smoluchowski-Poisson system}
\label{sec_brown}

We now consider the dynamics of a system of self-gravitating Brownian
particles in a space of dimension $D$. Like in Ref. \cite{charosi}, we
consider a high friction limit in order to simplify the problem. We thus 
study the Smoluchowski equation
\begin{equation}
{\partial\rho\over\partial t}=\nabla\biggl \lbrack
{1\over\xi}(T\nabla\rho+\rho\nabla\Phi)\biggr\rbrack,
\label{brown3}
\end{equation}
coupled to the Newton-Poisson equation (\ref{maxent4}). In the microcanonical ensemble, the temperature $T(t)$
evolves in time so as to satisfy the energy constraint
(\ref{maxent8}). In the canonical ensemble, the temperature is
constant. The Smoluchowski equation can be obtained from a variational
principle called the Maximum Entropy Production Principle
\cite{csr}. This variational approach is
interesting as it makes a direct link between the dynamics and the
thermodynamics. In the microcanonical description, the rate of
entropy production can be put in the form \cite{csr}
\begin{equation}
\dot S=\int {1\over T\rho\xi}(T\nabla\rho+\rho\nabla\Phi)^{2}d^{D}{\bf r}\ge 0.
\label{brown4}
\end{equation}
For a stationary solution, $\dot S=0$ and we obtain the Boltzmann
distribution (\ref{maxent10}) which is a critical point of
entropy. Considering a small perturbation around equilibrium, we can
establish the identity \cite{charosi}:
\begin{equation}
\delta^{2}\dot S=2\lambda\delta^{2}S\ge 0,
\label{brown5}
\end{equation}
where $\lambda$ is the growth rate of the perturbation defined such
that $\delta\rho\sim e^{\lambda t}$. This relation shows that a
stationary solution of the Smoluchowski-Poisson system is dynamically
stable if and only if it is a local entropy maximum. We get similar
results in the canonical ensemble with $J$ in place of $S$. The
relation (\ref{brown5}) has been found for other kinetic equations
satisfying a $H$-theorem (Chavanis, in preparation). In $D=2$,  an equation
morphologically similar to the Smoluchowski equation (\ref{brown3})
can been derived from the $N$-body Liouville equation for a gas of
point vortices \cite{kin}.

\subsection{Self-similar solutions of the Smoluchowski-Poisson system}
\label{sec_dim}

From now on, we set $M=R=G=\xi=1$. The equations of the problem become
\begin{equation}
{\partial\rho\over\partial t}=\nabla (T\nabla\rho+\rho\nabla\Phi),
\label{dim1}
\end{equation}
\begin{equation}
\Delta\Phi=S_{D}\rho,
\label{dim2}
\end{equation}
\begin{equation}
E={D\over 2}T+{1\over 2}\int \rho\Phi d^{D}{\bf r},
\label{dim3}
\end{equation}
with boundary conditions
\begin{equation}
{\partial\Phi\over\partial r}(0,t)=0, \qquad \Phi(1)={1\over 2-D},
\qquad T{\partial \rho\over\partial r}(1)+\rho(1)=0,
\label{dim4}
\end{equation}
for $D>2$. For $D=2$, we take $\Phi(1)=0$ on the boundary. Integrating
Eq.~(\ref{dim2}) once, we can rewrite the Smoluchowski-Poisson system
in the form of a single integrodifferential equation
\begin{equation}
{\partial\rho\over\partial t}={1\over r^{D-1}}{\partial\over\partial
r}\biggl\lbrace r^{D-1}\biggl (T{\partial\rho\over\partial
r}+{\rho\over r^{D-1}}\int_{0}^{r}\rho(r')S_{D}r^{'D-1}dr'\biggr
)\biggr \rbrace.
\label{dim5a}
\end{equation}
The Smoluchowski-Poisson system is also equivalent to a single
differential equation
\begin{equation}
\frac{\partial M}{\partial t}=T \left(\frac{\partial^2 M}{\partial r^2}
-\frac{D-1}{r}\frac{\partial M}{\partial r}\right)
+{1\over r^{(D-1)}}M\frac{\partial M}{\partial r},
\label{sca}
\end{equation}
for the quantity
\begin{equation}
M(r,t)=\int_{0}^{r}\rho(r')S_{D}r^{'D-1}dr',
\label{dint}
\end{equation}
which represents the mass contained within the sphere of radius
$r$. The appropriate boundary conditions are
\begin{equation}
M(0,t)=0,\qquad M(1,t)=1.
\label{dintb}
\end{equation}
It is also convenient to introduce the function $s(r,t)=M(r,t)/r^{D}$
satisfying
\begin{equation}
{\partial s\over\partial t}=T\biggl ({\partial^{2}s\over\partial
r^{2}}+{D+1\over r}{\partial s\over\partial r}\biggr )+\biggl
(r{\partial s\over\partial r}+Ds\biggr )s.
\label{seq}
\end{equation}

We look for self-similar solutions of the form
\begin{equation}
\rho(r,t)=\rho_{0}(t)f\biggl ({r\over r_{0}(t)}\biggr ), \qquad r_{0}=
\biggl ({T\over \rho_{0}}\biggr )^{1/2}.
\label{dim5}
\end{equation}
In terms of the mass profile, we have
\begin{equation}
M(r,t)=M_{0}(t)g\biggl ({r\over r_{0}(t)}\biggr ), \qquad {\rm
with}\qquad M_{0}(t)=\rho_{0}r_{0}^{D},
\label{dim6}
\end{equation}
and
\begin{equation}
g(x)=\int_{0}^{x}f(x')S_{D}x^{'D-1}dx'.
\label{dim7}
\end{equation}
In terms of the function $s$, we have
\begin{equation}
s(r,t)=\rho_{0}(t)S\biggl ({r\over r_{0}(t)}\biggr ), \qquad {\rm
with}\qquad S(x)={g(x)\over x^{D}}.
\label{dim6s}
\end{equation}

Substituting the {\it ansatz} (\ref{dim6s}) into Eq.~(\ref{seq}), we find that
\begin{equation}
{d\rho_{0}\over dt}S(x)-{\rho_{0}\over r_{0}}{dr_{0}\over dt}x
S'(x)={\rho_{0}^{2}}\biggl (S''(x)+{D+1\over
x}S'(x)+xS(x)S'(x)+DS(x)^{2}\biggr ),
\label{dim8}
\end{equation}
where we have set $x=r/r_{0}$. The variables of position and time
separate provided that there exists $\alpha$ such that
\begin{equation}
 \rho_{0}r_{0}^{\alpha}=\kappa,
\label{dim9}
\end{equation}
where $\kappa$ is a constant. In that case, Eq.~(\ref{dim8}) reduces to
\begin{equation}
{d\rho_{0}\over dt}\biggl (S(x)+{1\over\alpha}xS'(x)\biggr
)={\rho_{0}^{2}}\biggl (S''(x)+{D+1\over
x}S'(x)+xS(x)S'(x)+DS(x)^{2}\biggr ).
\label{dim10}
\end{equation}
Assuming that such a scaling exists implies that
$(1/\rho_{0}^{2})(d\rho_{0}/dt)$ is a constant that we arbitrarily set
equal to $\alpha$ (note that this convention is different from the one
adopted in Ref.~\cite{charosi}). This leads to
\begin{equation}
\rho_{0}(t)={1\over\alpha}(t_{coll}-t)^{-1},
\label{dim11}
\end{equation}
so that the central density becomes infinite in a finite time $t_{coll}$.
The scaling equation now reads
\begin{equation}
\alpha S+xS'=S''+{D+1\over x}S'+S(xS'+DS).
\label{scalingd}
\end{equation}
For $x\rightarrow +\infty$, we have asymptotically
\begin{equation}
S(x)\sim x^{-\alpha}, \qquad g(x)\sim x^{D-\alpha}, \qquad f(x)\sim
x^{-\alpha}.
\label{dim13}
\end{equation}

\subsection{Canonical ensemble}
\label{sec_Cane}

In the canonical ensemble in which the temperature $T$ is a constant,
we have \footnote{The case $T=0$ is treated in the Appendix
\ref{sec_cold}}
\begin{equation}
\alpha=2,\qquad \kappa=T.
\label{Cane1}
\end{equation}
In that case, the scaling equation (\ref{scalingd}) can be solved
analytically. Following a procedure similar to the one developed in
Ref.~\cite{charosi}, we find that
\begin{equation}
S(x)={4\over D-2+{x^{2}}}.
\label{Cane1a}
\end{equation}
Then, Eqs.~(\ref{dim6s}) and (\ref{dim7}) yield
\begin{equation}
g(x)={4x^{D}\over D-2+{x^{2}}},\qquad f(x)={4(D-2)\over
S_{D}}{D+x^{2}\over \bigl (D-2+{x^{2}}\bigr )^{2}}.
\label{Cane2a}
\end{equation}
According to Eqs.~(\ref{dim5}) and (\ref{dim11}), the central density evolves
with time like
\begin{equation}
\rho(0,t)=\rho_{0}(t)f(0)={2D\over (D-2)S_{D}}(t_{coll}-t)^{-1}.
\label{Cane2}
\end{equation}
According to Eqs.~(\ref{dim5}) and (\ref{dim6}), the core radius and
the core mass evolve like
\begin{equation}
r_{0}(t)=\sqrt{2T} (t_{coll}-t)^{1/2}, \qquad M_{0}(t)={1\over
2}(2T)^{D/2}(t_{coll}-t)^{{D\over 2}-1}.
\label{Cane3}
\end{equation}
Note that for $D>2$, the core mass goes to zero at the collapse time.
At $t=t_{coll}$, we get the singular profile
\begin{equation}
\rho(r,t=t_{coll})={4T(D-2)\over S_{D}r^{2}},\qquad M(r,t=t_{coll})=4Tr^{D-2}.
\label{Cane5}
\end{equation}

\subsection{Microcanonical ensemble}
\label{sec_Micro}

In the microcanonical ensemble, the exponent $\alpha$ is not
determined by simple dimensional analysis. In Ref.~\cite{charosi}, we
found numerically that the scaling equation (\ref{scalingd}) has physical
solutions only for $\alpha\le\alpha_{\rm max}$, with
$\alpha_{max}\simeq 2.21$ for $D=3$. We also argued that the system
will select the exponent $\alpha_{max}$, since it leads to a maximum
increase of entropy. In this section, we show that in the limit of
large dimension $D$, we can explicitly understand the occurrence of
such a $\alpha_{\rm max}$. In addition, we will present the derivation
of perturbative expansions for $\alpha_{\rm max}$ and the scaling
function $S$, in powers of $D^{-1}$.

Eq.~(\ref{scalingd}) can be formally integrated as a first order differential
equation (writing $S''=S'\times[S''/S']$), leading to an expression of
$S(x)$ as a function of $x$, $S(x)$ itself, and $S''(x)/S'(x)$:
\begin{equation}
\left|\frac{\alpha}{DS(x)}-1\right|=\left|\frac{\alpha}{DS(0)}-1\right|
\exp\left[\alpha\int_0^x\frac{y\,dy}{y^2(1-S(y))-y\frac{S''(y)}{S'(y)}-(D+1)}
\right].
\label{simp}
\end{equation}
We now define $x_0$, such that $S(x_0)=\alpha/D$. Since $S$ should be analytic,
the foregoing relation implies for $x\rightarrow x_{0}$,
\begin{equation}
\int_{0}^{x}{\alpha y\over F(y)}dy\sim \ln|x-x_{0}|,
\label{Micro1}
\end{equation}
where $F(y)$ is the function which occurs in the denominator of the
integral in Eq.~(\ref{simp}). From Eq.~(\ref{Micro1}), we must have
$F(y)=\alpha x_{0}(y-x_{0})$ for $y\rightarrow x_{0}$, which implies
$F(x_{0})=0$ and $F'(x_{0})=\alpha x_{0}$. These conditions can be
rewritten
\begin{equation}
x_{0}^{2}\biggl (1-{\alpha\over D}\biggr )-x_{0}{S''(x_{0})\over
S'(x_{0})}-(D+1)=0,
\label{condd1}
\end{equation}
\begin{equation}
(\alpha-2) x_0=-\frac{d}{dx}\left[x^2S(x)+x
\frac{S''(x)}{S'(x)}\right]_{x_0}.
\label{condd2}
\end{equation}
This preparatory work now allows the introduction of a systematic expansion in
large dimension $D$ for the scaling function $S$, the scaling exponent
$\alpha$, and $x_0$. In this limit, let us neglect the contribution of
the terms which are not of order $D$ in the right-hand side of
Eq.~(\ref{scalingd}). This actually amounts to take $F(y)\simeq y^2-D$ in
Eq.~(\ref{simp}). Within this approximation, we find
\begin{equation}
\left|\frac{\alpha}{DS(x)}-1\right|=\left|\frac{\alpha}{DS(0)}-1\right|
\times \left|\frac{x^2}{D}-1\right|^{\alpha/2},
\label{simpinf}
\end{equation}
which is an analytic function only if $\alpha=2$. This leads to
$x_0=\sqrt{D}$, and to the more explicit form for $S$,
\begin{equation}
S(x)=\frac{S(0)}{1+\left(\frac{DS(0)}{2}-1\right)\frac{x^2}{D}}.
\label{sinf}
\end{equation}
$S(0)$ remains undetermined, and will be fixed by the next order
approximation.  Indeed, we can iteratively solve the full scaling
equation Eq.~(\ref{scalingd}), by reinserting the zeroth order
solution into Eq.~(\ref{simp}), and eventually continue this process
with the new improved scaling function, and so forth... Thus, expressing
the conditions of Eq.~(\ref{condd1}) and Eq.~(\ref{condd2}), and defining
$z=\frac{DS(0)}{2}$ (which will be of order $O(1)$), we obtain
\begin{equation}
x_0^2=D+\frac{4}{z}+O(D^{-1}), {\rm \ or\ }
x_0=\sqrt{D}\left(1+\frac{2}{zD}+O(D^{-2})\right),
\label{x0d}
\end{equation}
and
\begin{equation}
\alpha-2=\frac{4}{D}\left[\frac{1}{z}-\frac{2}{z^2}\right]+O(D^{-2}).
\label{alphad1}
\end{equation}
Eq.~(\ref{alphad1}) provides a relation between the possible values
for $\alpha$ and the associated value of $S(0)=\frac{2z}{D}$. Note
that the function of $z$ in the right-hand side of Eq.~(\ref{alphad1})
has a well defined maximum. Hence, up to order $O(D^{-1})$, we find
that $\alpha\leq \alpha_{\rm max}$, with
\begin{equation}
\alpha_{\rm max}=2+\frac{1}{2}\,D^{-1}+O(D^{-2}),
\label{alphadexp1}
\end{equation}
which is associated to the value $z=4+O(D^{-1})$ or
$S(0)=\frac{8}{D}+O(D^{-2})$. As $\alpha$ is necessarily greater than
2 (as the temperature cannot vanish), a solution exists for any
$\alpha\in[2,\alpha_{\rm max}]$. As already mentioned, $\alpha_{\rm
max}$ is dynamically selected as it leads to the maximum divergence of
the entropy and the temperature (see Eq.~(\ref{atheta}) below).

Inserting Eq.~(\ref{sinf}) into
Eq.~(\ref{simp}), we find the next order approximation for $S$
\begin{equation}
\left|\frac{\alpha}{DS(x)}-1\right|=\left|\frac{\alpha}{DS(0)}-1\right|
\times \left|\frac{x^2}{x_0^2}-1\right|^{\frac{\alpha}{2}(1-\phi)}\times
 \left[\frac{x^2}{x_1^2}+1\right]^{\frac{\alpha\phi}{2}},
\label{snext}
\end{equation}
where $x_0$ is given by Eq.~(\ref{x0d}), and $x_1$ and $\phi$ are defined
by
\begin{equation}
x_1^2=\frac{D}{z-1}+\frac{2(z-2)}{z(z-1)}+O(D^{-1}),\quad
\phi=\frac{2}{D}\left[\frac{1}{z}-\frac{2}{z^2}\right]+O(D^{-2}).
\label{phix1}
\end{equation}
Again, the analyticity condition imposes that
$\frac{\alpha}{2}(1-\phi)=1$, which exactly leads to Eq.~(\ref{alphad1}),
and to the following explicit form for $S$:
\begin{equation}
S(x)=\frac{\alpha}{D}\left[1+\left(1-\frac{\alpha}{2z}\right)
\left(\frac{x^2}{x_0^2}-1\right)
 \left(\frac{x^2}{x_1^2}+1\right)^{\frac{\alpha}{2}-1}\right]^{-1}.
\label{snextexp}
\end{equation}
This improved scaling function can be inserted again into the
conditions expressed by Eq.~(\ref{condd1}) and Eq.~(\ref{condd2}),
leading to the next order term in the expansion of $\alpha$. After
elementary but cumbersome calculations, we end up with
\begin{equation}
\alpha-2=\frac{4}{D}\left[\frac{1}{z}-\frac{2}{z^2}\right]
+\frac{8}{D^2}\left[\frac{5}{z}-\frac{26}{z^2}+
\frac{31}{z^3}-\frac{6}{z^4}
-\left(\frac{1}{z}-\frac{7}{z^2}+ \frac{14}{z^3}-\frac{8}{z^4}\right)\ln z
\right]+ O(D^{-3}).
\label{alphad2}
\end{equation}
This function has again a well defined maximum for
\begin{equation}
z=\frac{D}{2}S(0)=4+\left(\frac{41}{2}-6\ln 2\right)D^{-1}+ O(D^{-2}),
\label{zd2}
\end{equation}
associated to the value
\begin{equation}
\alpha_{\rm max}=2+\frac{1}{2}\,D^{-1}+\frac{11}{16}\,D^{-2}+O(D^{-3}).
\label{alphaexp2}
\end{equation}
This expansion gives $\alpha_{\rm max}=2.24...$ in $D=3$, in fair
agreement with the exact value $\alpha_{\rm max}=2.2097...$ obtained
numerically in \cite{charosi}. In addition, the exponent $\alpha=2$ is
associated to $z=2+4D^{-1}+ O(D^{-2})$.  In principle, these
expansions can be systematically pursued to the prize of increasingly
complicated calculations.

Finally note that Eqs.~(\ref{dim5}) and (\ref{dim11}) lead to the
following exact asymptotic for the central density $\rho(0,t)$:
\begin{equation}
\rho(0,t)\sim {K_D(\alpha)}({t_{coll}-t})^{-1},\quad
K_D(\alpha)=\frac{2z(\alpha)}{\alpha S_D},
\end{equation}
where we have used $f(0)=DS(0)/S_{D}$ and the definition of $z$. The
function $z(\alpha)$ is determined implicitly by Eq.~(\ref{alphad2}),
up to order $O(D^{-2})$. For the special cases $\alpha=2$ and
$\alpha=\alpha_{\rm max}$, we respectively find
\begin{eqnarray}
K_D(2)&=& 2S_D^{-1}\left(1+2D^{-1}+O(D^{-2})\right), \\
K_D(\alpha_{\rm max})&=& 4S_D^{-1}\left(1+\left(\frac{39}{8}-
\frac{3}{2}\ln 2\right)D^{-1}+O(D^{-2})\right),
\end{eqnarray}
which shows that $K_D(\alpha_{\rm max})$ is substantially greater than
$K_D(2)$ (twice bigger in the infinite $D$ limit, the ratio being even
bigger for finite $D$, as $\frac{39}{8}- \frac{3}{2}\ln 2\approx
3.835...>2$). This substantial difference was noted in
Ref.~\cite{charosi}, in the case $D=3$. Finally, as expected in the
microcanonical ensemble, the temperature diverges during the collapse
as $T(t)\sim (t_{coll}-t)^{-\alpha_T}$ with $\alpha_{T}=1-2/\alpha$;
see Eqs.~(\ref{dim5})(\ref{dim9})(\ref{dim11}).  The strongest
divergence is obtained for $\alpha=\alpha_{\rm max}$. According to
Eq.~(\ref{alphaexp2}), we have
\begin{equation}
\alpha_T=2-\frac{2}{\alpha_{\rm max}}=
\frac{1}{4}\,D^{-1}+\frac{9}{32}\,D^{-2}
+O(D^{-3}).\label{atheta}
\end{equation}
If we plug $D=3$ in Eq.~(\ref{atheta}), we find the estimate
$\alpha_T\approx 0.11...$ in fair agreement with the exponent measured
numerically in \cite{charosi}, $\alpha_T\approx 0.1$.

\section{The two-dimensional case}
\label{sec_twoD}

\subsection{The critical temperature}
\label{sec_crit}

In two dimensions, the dynamical equation (\ref{sca}) for the mass
profile reads
\begin{equation}
\frac{\partial M}{\partial t}=4T u\frac{\partial^2 M}{\partial u^2}
+2M\frac{\partial M}{\partial u},
\label{dyn2d}
\end{equation}
after the change of variable $u=r^2$ has been effected. Looking for a
stationary solution, and using $uM''=(uM')'-M'$, Eq.~(\ref{dyn2d}) is
readily integrated leading to
\begin{equation}
M(u)={4T\over 4T-1}\ \frac{u}{1+\frac{u}{4T-1}},
\label{stat2d}
\end{equation}
Note that $M(1)=1$, ensuring that the whole mass is included in this
solution. Using $\rho=M'/\pi$, we find that the density profile is
given by
\begin{equation}
\rho(r)=\frac{4\rho_0}{\pi}\frac{1}{(1+(r/r_0)^2)^{2}},
\label{rhost2d}
\end{equation}
with
\begin{equation}
 r_0=\sqrt{4T-1}\qquad {\rm and}\qquad \rho_0 r_0^{2}=T.
\label{r0stat}
\end{equation}
This solution exists provided that $T>T_c=1/4$, which defines the
collapse temperature. We have thus recovered the result
(\ref{thermo11}) by a slightly different method. Note that the value
of $T_{c}$ and the dependance of $r_{0}$ and $\rho_{0}$ with the
temperature coincide with the exact results obtained within conformal
field theory \cite{abdalla}. In the following, $T$ is set constant as
we have already seen that the gravothermal does not exist in the
microcanonical ensemble in two dimensions.

\subsection{Scaling collapse for $T=T_c$}

We now address the dynamics at the critical temperature
$T=T_c=1/4$. We note that contrary to what happens in other
dimensions, the central density diverges at $T_c$. Thus, in analogy
with critical phenomena, we anticipate a scaling solution for
$M(u,t)$, of the form
\begin{equation}
M(u,t)\approx\frac{(a(t)+1)u}{1+a(t)u},
\label{tc2d}
\end{equation}
which preserves the scaling form obtained above $T_c$, and which
satisfies the boundary condition $M(1,t)=1$. The corresponding density
profile is
\begin{equation}
\rho(r,t)=\frac{a(t)+1}{\pi}{1\over (1+a(t)r^2)^{2}}.
\label{scarho2d}
\end{equation}
The central density
\begin{equation}
\rho(0,t)=\frac{a(t)+1}{\pi},
\label{rhoa}
\end{equation}
is expected to diverge for $t\rightarrow +\infty$, so that $a(t)$ is
also expected to diverge.

Inserting the {\it ansatz} Eq.~(\ref{tc2d}) into Eq.~(\ref{dyn2d})
shows that the left-hand term is indeed negligible compared to both
terms of the right-hand side, to leading order in $a$. So far, this
prevents us from determining a dynamical equation for $a$. In order to
achieve that, we must solve Eq.~(\ref{dyn2d}) to the next order in
$a^{-1}$. We thus look for a solution of the form
\begin{equation}
M(u,t)=\frac{a(t)u}{1+a(t)u}+a(t)^{-1}h(u,t),
\end{equation}
where $h(u,t)$ is expected to be of order $O(1)$, and should satisfy
$h(0,t)=0$ and $h(1,t)=1$ (the total integrated mass should be 0 and
1, respectively, for $u=0$ and $u=1$), and $\frac{\partial h}{\partial
u}(0,t)=0$, which ensures that Eq.~(\ref{rhoa}) is exactly obeyed,
defining $a(t)$ without any ambiguity. The contribution of
$\frac{\partial M}{\partial t}$ in the left-hand side of
Eq.~(\ref{dyn2d}) is dominated by the time derivative of
Eq.~(\ref{tc2d})~:
\begin{equation}
\frac{\partial }{\partial t}\left[\frac{(1+a(t))u}{1+a(t)u}\right]=
\frac{u(1-u)}{(1+au)^2}\frac{da}{dt},
\end{equation}
which will be checked self-consistently hereafter. In addition, non
linear terms in $h$ in the right-hand side are also
negligible. Therefore, $h$ satisfies
\begin{equation}
\frac{au(1-u)}{(1+au)^2}\frac{da}{dt}=u\frac{\partial^2 h}{\partial u^2}
+2\frac{\partial }{\partial u}\left(\frac{au}{1+au}h\right).
\label{eqh}
\end{equation}
Actually, for a given time, this equation becomes an ordinary
differential equation involving only one variable $u$, as $a$ and
${da}/{dt}$ appear as parameters. Eq.~(\ref{eqh}) can be integrated
leading to a first order equation in $h$, which can be easily
solved. Defining $v=au$, we finally get
\begin{eqnarray}
&&h(u,t)=a^{-1}\left(1+\frac{2}{a}\right)\frac{da}{dt}(1+v)^{-2}
\times\label{h}\\
&&\left[(v^2-1)\ln(1+v)+v(1-2v)+2v
\int_0^v\frac{\ln(1+z)}{z}\,dz-\frac{2v^2+v^3}{2(a+2)}\right],
 \nonumber
\end{eqnarray}
which depends on time only through the variable $a$ and ${da}/{dt}$.
Now, ${da}/{dt}$ is determined by imposing the boundary condition
$h(1,t)=1$, which leads to
\begin{equation}
\frac{da}{dt}=\frac{a}{\ln a-5/2}\left[1+O(\ln a^{-2})\right].
\label{dadt}
\end{equation}
One can solve iteratively Eq.~(\ref{dyn2d}), by adding the time
derivative of the above solution to the left-hand side, in order to
compute an improved $h$. To leading order, the solution of
Eq.~(\ref{h}) is preserved. However, new terms are generated which are
important for $v\sim a$ ($u\sim 1$), and which generate terms of order
$O(a/\ln^3a)$ is the expansion for ${da}/{dt}$. This explains the form
of the error term in Eq.~(\ref{dadt}).

\begin{figure}
\centerline{
\psfig{figure=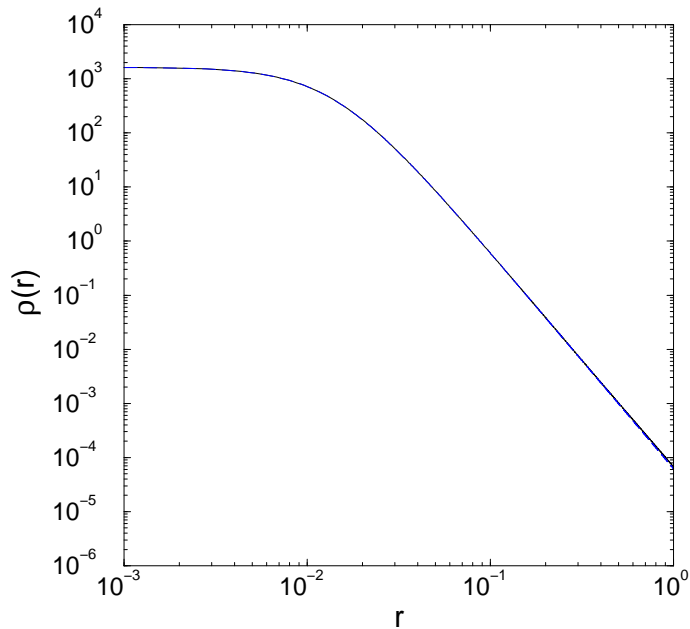,angle=0,height=8.5cm}}
\caption{At $T=T_c=1/4$, and when the central density
has reached the value $\rho(0,t)\approx 1644.8...=\frac{a(t)+1}{\pi}$
($a(t)\approx 5166.3...$), we have plotted the result of the numerical
calculation compared to our exact scaling form
$\rho(r,t)=\frac{a(t)+1}{\pi}(1+a(t)r^2)^{-2}$ obtained in
Eq.~(\ref{scarho2d}). The two curves are indistinguishable as the
relative error is, as predicted, of order $a^{-1}\sim 10^{-4}$.  Note
finally that for this range of density, the density contrast is huge,
of order $10^7$.}
\label{colld_fig1}
\end{figure}

Integrating Eq.~(\ref{dadt}) for large time, we get the exact
asymptotic expansion for large time
\begin{equation}
a(t)=\exp\left(\frac{5}{2}+\sqrt{2t}\right)\left[1+O(t^{-1/2}\ln
t)\right].
\label{at}
\end{equation}
For $t\rightarrow +\infty$, the central density diverges like $a(t)$
and the core radius goes to zero like $a(t)^{-1/2}$. In addition, the
scaling solution (\ref{scarho2d}) at $T=T_{c}$ goes to a Dirac peak
containing the whole mass (see Eq.~(\ref{thermo12})), as the decay
exponent of the scaling function is 4, which is strictly greater than
2.

\begin{figure}
\centerline{
\psfig{figure=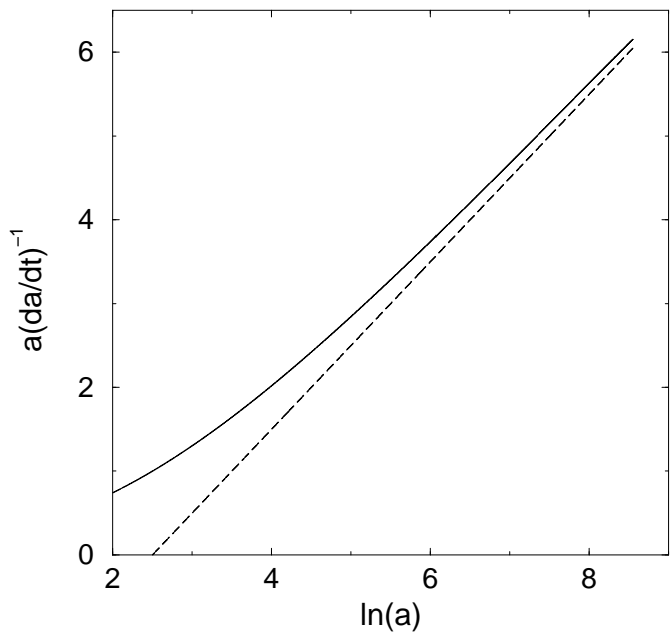,angle=0,height=8.5cm}}
\caption{ We plot $a(da/dt)^{-1}$ as a function of $\ln a$, which is
predicted to behave as $a(da/dt)^{-1}\sim
\ln a-5/2+O([\ln a]^{-1})$ (see Eq.~(\ref{dadt})).
Even for the moderate range of accessible densities ($a_{\rm max}\sim
5166$), we clearly find that the numerical result evolves toward the
theoretical asymptotic (dashed line). }
\label{colld_fig2}
\end{figure}

\subsection{Collapse for $T<T_c$}
\label{sec_TinfTc}

For $D=2$, the scaling equation associated to Eq.~(\ref{sca}) does not
display any physical solution when numerically solved. In this
section, we thus present a special treatment adapted to this case.
The principal difference with other dimensions is the divergence of
the central density at $T_c$, and the occurrence of a scaling solution
at this temperature.

Strictly below $T_c$, we expect a finite time collapse. Close to the
center, the solution takes the form
\begin{equation}
M(u,t)\approx 4T\frac{a(t)u}{1+a(t)u},
\label{col2d0}
\end{equation}
where again the left-hand side of Eq.~(\ref{dyn2d}) is negligible
compared to each term on the right-hand side. We thus look for a
solution of the type
\begin{equation}
M(u,t)= 4T\frac{a(t)u}{1+a(t)u}+h(u,t),
\label{col2d}
\end{equation}
where $h$ is now of order $O(1)$ as it contains a finite fraction of
the total mass, since the first term contains a mass of order
$4T<1$. Inserting this {\it ansatz} in the dynamical equation
Eq.~(\ref{dyn2d}), we obtain
\begin{equation}
\frac{1}{4T}\frac{\partial h}{\partial t}+\frac{da}{dt}
\frac{u}{(1+au)^2}=
u\frac{\partial^2 h}{\partial u^2} +2\frac{\partial }{\partial
u}\left(\frac{au}{1+au}h\right)+2h\frac{\partial h}{\partial u}.
\label{ew}
\end{equation}
One can look for a scaling solution of the type
\begin{equation}
h(u,t)=a^{\gamma-1}H(au), {\rm \ with \ } H(v)\sim cv^{1-\gamma},
{\rm\ when\ }v\to+\infty,
\label{asymf}
\end{equation}
so that the mass included in this scaling profile $h(1,t)=c=O(1)$.
With this definition, the density profile decays for large distance as
$\rho\sim r^{-\alpha}$, with $\alpha=2\gamma$.  Inserting this {\it
ansatz} in Eq.~(\ref{ew}), we obtain
\begin{equation}
\left[\frac{1}{4T}\left(vH'+(\gamma-1)H\right)+a^{1-\gamma}
\frac{v}{(1+v)^2}\right]\frac{da}{dt}a^{-2}=vH''+2
\left(\frac{v}{1+v}H\right)'+
2a^{\gamma-1}H H',
\label{scacol2d}
\end{equation}
where derivatives are with respect to the variable $v$. We are free to
choose $a(t)=\pi\rho(0,t)/(4T)$, so that $H'(0)=H(0)=0$. For small
$v$, Eq.~(\ref{scacol2d}) leads to
\begin{equation}
\frac{da}{dt}=H''(0)a^{\gamma+1}.
\label{dadtg}
\end{equation}
Eq.~(\ref{scacol2d}) has a global scaling solution only for
$\gamma=1$.  However, we know that in this case the scaling equation
obtained by setting $\gamma=1$ does not display any physical
solution. Thus, we conclude that there is no scaling solution obtained
by imposing that all terms in Eq.~(\ref{scacol2d}) scale the same
way. However, as we will see in the section devoted to numerical
simulations, the direct simulation of Eq.~(\ref{dyn2d}) seems to
display a scaling solution with $\gamma\approx 0.6-0.7$ for
numerically accessible densities. Strictly speaking, this is totally
excluded by the above equation, except if one allows $\gamma$ to very
slowly depend on the density or $a$. For a given $a$, we thus want to
solve Eq.~(\ref{scacol2d}), where the boundary conditions will
ultimately select the effective value of $\gamma$, and that of
${da}/{dt}$. More precisely, once we impose $H'(0)=H(0)=0$, and the
condition of Eq.~(\ref{dadtg}), we end up with a shooting problem for
$H''(0)$ and $\gamma$. For large $a$, and $v\ll a$, it is clear that
the non linear term of the right-hand side of Eq.~(\ref{scacol2d})
becomes irrelevant, and we drop it from now on.

In order to understand the origin of this shooting problem, and to
obtain an accurate estimate of $\gamma$, let us solve
Eq.~(\ref{scacol2d}) in the limit of very large $a$, in the range
$1\ll v\ll a$. In this regime, Eq.~(\ref{scacol2d}) simplifies to the
following equation
\begin{equation}
\left[\frac{1}{4T}\left(vH'+(\gamma-1)H\right)+
a^{1-\gamma}v^{-1}\right]\omega=vH''+2H',
\label{sca2dsimp}
\end{equation}
where
\begin{equation}
\omega=\frac{da}{dt}a^{-2}=H''(0)a^{\gamma-1}.
\label{omega}
\end{equation}
Let us now multiply this equation by $v^{\gamma-2}$ and integrate the
resulting equation. After elementary manipulations, we obtain
\begin{equation}
H'+\left[\frac{3-\gamma}{v}-\frac{\omega}{4T}\right]H=-\frac{\omega
c}{4T} v^{1-\gamma}-
\frac{\omega a^{1-\gamma}}{2-\gamma}v^{-1}+(2-\gamma)(3-\gamma)v^{1-\gamma}
\int_v^{+\infty}w^{\gamma-3}H(w)\,dw,\label{eqf}
\end{equation}
where $c\sim O(1)$, which has been defined in Eq.~(\ref{asymf}),
appears here as an integration constant.  Then, one can integrate this
differential equation which leads to the following self-consistent
relation for $H$:
\begin{equation}
H(v)=v^{\gamma-3}\exp\left(\frac{\omega v}{4T}\right)
\int_v^{+\infty}w^{3-\gamma}\exp\left(-\frac{\omega w}{4T}\right)F(w)\,dw,
\label{eqff}
\end{equation}
where $F$ is defined as the opposite of the right-hand side of
Eq.~(\ref{eqf}):
\begin{equation}
F(v)=\frac{\omega c}{4T} v^{1-\gamma}+
\frac{\omega a^{1-\gamma}}{2-\gamma}v^{-1}-(2-\gamma)(3-\gamma)v^{1-\gamma}
\int_v^{+\infty}w^{\gamma-3}H(w)\,dw.
\end{equation}
Eq.~(\ref{sca2dsimp}) implies that $H(v)\sim\ln v$, when $v\to 0$ (of
course, this apparent divergence does not occur in the full dynamical
equation Eq.~(\ref{scacol2d})). Considering the prefactor
$v^{\gamma-3}$ in Eq.~(\ref{eqff}), this behavior can be obtained if
and only if
\begin{equation}
\int_0^{+\infty}w^{3-\gamma}\exp\left(-\frac{\omega w}{4T}\right)
F(w)\,dw=0.\label{cond1}
\end{equation}
As $\omega$ is expected to go to zero for large $a$ as $\gamma<1$ (see
Eq.~(\ref{omega})), the dominant contribution of the integral of the
third term in the definition of $F$ comes from the large $w$ region,
for which $H$ can be replaced by its asymptotic form. Hence, defining
$\Gamma(x)=\int_0^{+\infty}w^x\exp(-w)\,dw$ and
$\varepsilon=1-\gamma$, and using Eq.~(\ref{omega}), the condition
expressed in Eq.~(\ref{cond1}) can be rewritten
\begin{equation}
c=\frac{\Gamma(1+\varepsilon)}{\varepsilon(1+\varepsilon)^2
\Gamma(1+2\varepsilon)}H''(0)\left(\frac{H''(0)}{4T}\right)^\varepsilon
a^{-\varepsilon^2}.
\end{equation}
As $c$ is of order $O(1)$, we find that $\varepsilon\to 0$ as
$a\to+\infty$. More precisely, in this limit, $\varepsilon$ is the
solution of the following implicit equation
\begin{equation}
\varepsilon=\sqrt{\frac{\ln(K/\varepsilon)}{\ln a}},
\label{impe}
\end{equation}
where $K=H''(0)/c+O(\varepsilon)$. Finally, we obtain
\begin{equation}
\varepsilon=1-\gamma=\sqrt{\frac{\ln\ln a}{2\ln a}}
\left(1+O([\ln\ln a]^{-1})\right).
\label{expe}
\end{equation}
In conclusion, although the solution is not strictly speaking a true
scaling solution, the explicit dependence of $\gamma$ on $a$ is so
weak that an apparent scaling should be seen with an effective
$\gamma$ almost constant for a wide range of values of $a$. Hence, the
total density profile is the sum of the scaling profile obtained at
$T_c$ with a $T/T_c$ weight (behaving as a Dirac peak of weight
$T/T_c$, at $t=t_{coll}$) and of a pseudo-scaling solution associated
to an effective scaling exponent slowly converging to $\alpha=2$.

Let us illustrate quantitatively the time dependence of
$\alpha=2\gamma$. For example, taking arbitrarily $K=1$ (the
dependence on $K$ is weak and vanishes for large $a$),
Eq.~(\ref{impe}) and Eq.~(\ref{expe}) respectively lead to
$\gamma(a=10^3)=0.624...$ and $\gamma(a=10^3)=0.626...$, and to
$\gamma(a=10^5)=0.684...$ and $\gamma(a=10^5)=0.674...$ (note that the
error between the asymptotic expansion of Eq.~(\ref{expe}) and the
implicit expression first grows before slowly decaying for $a\gg
10^{12}$~!). Finally, for the maximum value of $a$ accessible
numerically of order $a\sim 10^4$, we expect to observe an apparent
scaling solution with $\gamma\approx 0.65$, or $\alpha=2\gamma\approx
1.3$.
\vskip 1cm

\begin{figure}
\centerline{
\psfig{figure=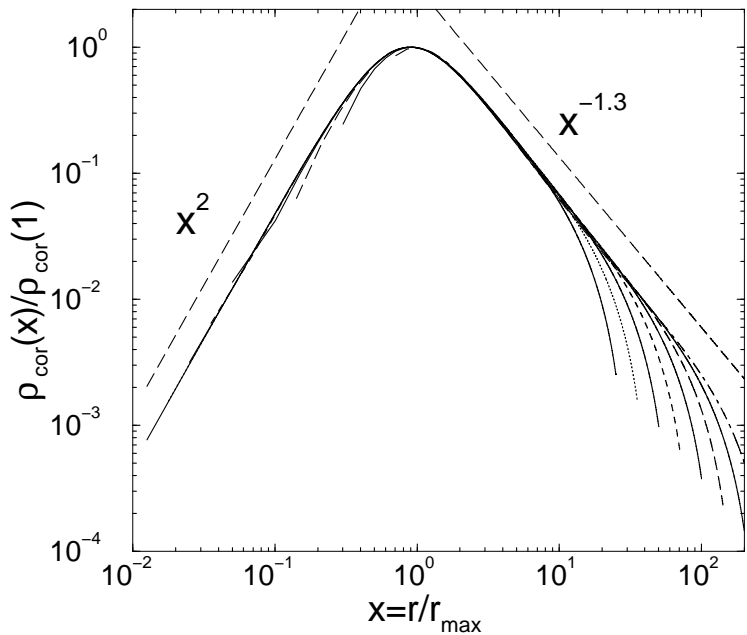,angle=0,height=8.5cm}}
\caption{At $T=T_{c}/2=1/8$, we have extracted the next
correction to scaling $\rho_{\rm cor}=\rho-4T\rho_{T=T_c}$, where
$\rho_{T=T_c}$ is defined in Eq.~(\ref{scarho2d}). We have then
plotted $\rho_{\rm cor}(r,t)/\rho_{\rm cor}(r_{\rm max}(t),t)$ as a
function of $x=r/r_{\rm max}(t)$, where $r_{\rm max}(t)$ is defined as
the location of the maximum of $\rho_{\rm cor}(r,t)$. Consistently
with the apparent scaling observed, we found $r_{\rm
max}^{-1}(t)\sim\sqrt{a}\sim\sqrt{\rho_{\rm cor}(r_{\rm
max}(t),t)}$. For $a=2^{n-1}\times 100$ ($n=1,...,8$), we have
obtained a convincing data collapse associated to
$\alpha=2\gamma\approx 1.3$, in agreement with the theoretical
estimate of $\gamma$, in this range of $a$. }
\label{colld_fig3}
\end{figure}

\subsection{Numerical simulations}

In this subsection, we present numerical simulations concerning the
two-dimensional case. Indeed, the three-dimensional case has been
extensively studied in \cite{charosi}.  It has been shown that the
scaling function as well as the corrections to scaling (which have
been calculated for the canonical ensemble in
\cite{charosi}) are perfectly described by the theory. In addition, as
the system behaves qualitatively the same for any dimension $D>2$ in
both thermodynamical ensembles, we have decided to focus on the
numerical study of the $D=2$ case, which displays some very rich
behaviors, as exemplified in the present section.

We consider the system in the canonical ensemble, as the gravitational
collapse does not occur in the microcanonical ensemble. In Fig.
\ref{colld_fig1}, we
show the scaling function at $T_c$, as given by Eq.~(\ref{scarho2d}),
finding a perfect agreement.  In Fig.~\ref{colld_fig2}, we also
display $a(da/dt)^{-1}$ as a function of $\ln a$, and find an
asymptotic behavior fully compatible with that given by
Eq.~(\ref{dadt}).

\begin{figure}
\centerline{
\psfig{figure=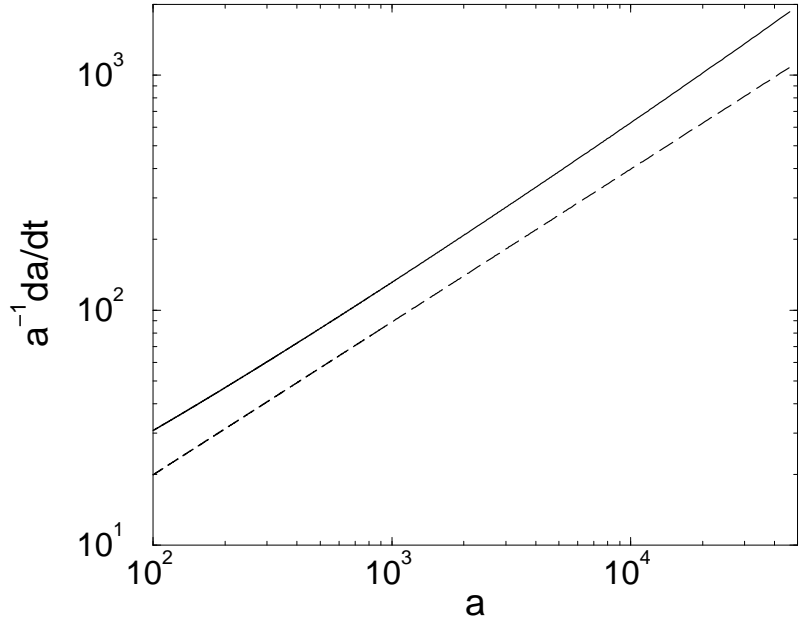,angle=0,height=8.5cm}}
\caption{We plot $a^{-1}(da/dt)\sim a^{\gamma}$ as a function
of $a$, in order to extract the effective value of $\gamma$ directly
from the time evolution of the central density. We find that the
effective $\gamma$ is slowly growing with time, as predicted, and is
of order $\gamma=\alpha/2\approx 0.65$ (the dashed line has a slope
equal to 0.65), which is fully compatible with the value extracted
from Fig.~\ref{colld_fig3}, and the value expected from
Eq.~(\ref{impe}) in this range of $a$. }
\label{colld_fig4}
\end{figure}

Below $T_c$, and in the accessible range of $a$ (up to $a\sim 10^5$),
we find an apparent scaling regime with $\alpha=2\gamma\approx 1.3$,
as predicted in Sec. \ref{sec_TinfTc}.  This is illustrated in
Fig.~\ref{colld_fig3}, for $T=T_{c}/2=1/8$. Note that the effective
$\gamma$ or $\alpha$ can also be extracted from the time evolution of
$a(t)$ or the central density (see Eq.~(\ref{dadtg})).  In
Fig.~\ref{colld_fig4}, we show that this way of measuring $\gamma$ is
fully compatible with the value of the effective exponents
$\alpha=2\gamma\approx 1.3$

\section{The one dimensional case}
\label{sec_oneD}

When an equilibrium state exists, there is little hope to be able to
solve the full Smoluchowski-Poisson system analytically in order to
study the relaxation towards equilibrium. We shall consider a simpler
problem in which a test particle evolves in a medium of field
particles at statistical equilibrium (thermal bath approximation). The
particles are assumed to create a stationary potential $\Phi_{eq}({\bf r})$ which
induces a drift of the test particle along the gradient of
$\Phi_{eq}$. In addition, the test particle is assumed to experience a
diffusion process.  If $\rho$ denotes the density probability of
finding the test particle in ${\bf r}$ at time $t$, we expect the
evolution of $\rho$ to be determined by a Smoluchowski equation of the
form
\begin{equation}
{\partial\rho\over\partial t}=\nabla(T\nabla\rho+\rho\nabla\Phi_{eq}),
\label{schro1}
\end{equation}
where $\Phi_{eq}({\bf r})$ is solution of the Boltzmann-Poisson
equation (\ref{maxent11}).  This means that we replace the true
potential by its equilibrium value but still allow the density $\rho$
to vary with time.  As we shall see, it is possible to solve the
Smoluchowski equation (\ref{schro1}) analytically in $D=1$ by using an
analogy with a problem of quantum mechanics. An equation of the form
(\ref{schro1}) has been proposed in Refs. \cite{drift,kin} to model
the motion of a test vortex in a bath of field vortices at statistical
equilibrium. In that context, Eq. (\ref{schro1}) can be formally
derived from the $N$-body Liouville equation of the system by using projection
operator technics.

It is well-known that a Fokker-Planck equation like Eq. (\ref{schro1})
can be formally transformed in\-to a Schr\"o\-din\-ger equation with
imaginary time. Indeed, performing the change of variable
\begin{equation}
\rho=\psi e^{-{1\over 2T}\Phi_{eq}},
\label{schro2}
\end{equation}
we find that the evolution of $\psi$ is determined by an equation of
the form
\begin{equation}
{\partial\psi\over\partial t}=T\Delta\psi+\biggl ({1\over
2}\Delta\Phi_{eq}-{1\over 4T}(\nabla\Phi_{eq})^{2}\biggr )\psi.
\label{schro3}
\end{equation}
This can be written as a Schr\"odinger-type equation
\begin{equation}
{\partial\psi\over\partial t}=T\Delta\psi-V({\bf r})\psi,
\label{schro3b}
\end{equation}
with a potential $V({\bf r})=-{1\over 2}\Delta\Phi_{eq}+{1\over
4T}(\nabla\Phi_{eq})^{2}$. So far, this transformation is
general. If we now consider the one-dimensional case, the
Boltzmann-Poisson equation (\ref{schro1}) can be solved
analytically and the potential $V(r)$ determined explicitly.
Introducing the notations $\xi=\alpha r/\sqrt{2}R$ and
$\tau=\alpha^{2}Tt/2R^{2}$ and using Eq. (\ref{emden13}) we can
rewrite Eq.~(\ref{schro3}) in the form
\begin{equation}
{\partial\psi\over\partial \tau}=\Delta\psi+\biggl ({2\over
\cosh^{2}\xi}-1\biggr )\psi.
\label{schro5}
\end{equation}
A separation of the variables can be effected by the substitution
\begin{equation}
\psi(\xi,t)=\phi(\xi)e^{-\lambda t} \qquad (\lambda\ge 0),
\label{schro6}
\end{equation}
where $\phi$ is solution of the ordinary differential equation
\begin{equation}
{d^{2}\phi\over d\xi^{2}}+2\biggl (E+{1\over \cosh^{2}\xi}\biggr
)\phi=0,
\label{schro7}
\end{equation}
where we have set $\lambda-1=2E$. The solutions of this
Schr\"odinger equation are described in detail in
Ref.~\cite{landau}. The spectrum of positive energies is
continuous. The spectrum of negative energies is discrete and
reduces to $E_{0}=-1/2$ (fundamental state). The first excited
state in the continuum is $E_{1}=0$. We can check that the
corresponding eigenfunctions are $\phi_{0}=1/\cosh\xi$ and
$\phi_{1}=\tanh\xi$. In order to obtain the qualitative behavior
of the time dependent solution of Eq. (\ref{schro1}), we neglect
the contribution from the continuum states with $E>0$, only
keeping the $E=-1/2$ and $E=0$ eigenstates.

Within this approximation and for sufficiently large times, we obtain
\begin{equation}
\psi(\xi,\tau)={A\over\cosh\xi}+B\tanh\xi\ e^{-\tau}, \qquad
(\tau\rightarrow +\infty).
\label{schro8}
\end{equation}
where $A$ and $B$ are constant. Returning to original variables, we
get
\begin{equation}
\rho(r,t)=\rho_{eq}(r)\biggl \lbrace 1+C\sinh\Bigl ({\alpha r\over\sqrt{2}R}
\Bigr )e^{-{\alpha^{2}T\over 2R^{2}}t}+...\biggr\rbrace,
\label{schro9}
\end{equation}
where $\rho_{eq}$ is given by Eq.~(\ref{thermo14}) and $C=B/A$ is a
constant. We find that the relaxation time is given by $t_{relax}=
2R^{2}/\alpha^{2}T$.

\section{Conclusion}

In this paper, we have studied the Boltzmann-Poisson equation and
the Smoluchowski-Poisson system in various dimensions of space.
Our study shows in particular how the nature of the problem
changes as we pass from $D=3$ to $D=2$. We showed that the
dimension $D=2$ is critical in the sense that the results
obtained for $D>2$ diverge if they are naively extrapolated to
$D=2$. On a physical point of view, the two-dimensional problem
differs from the $D>2$ case in two respects: in the 2D case, the
central density of the equilibrium state is infinite at the
critical temperature $T_{c}$ while it is finite at $T_{c}$ in
higher dimensions. On the other hand, in $D=2$, the self-similar
collapse results in a Dirac peak which contains a finite fraction
of mass, while for $D>2$, the mass contained in the core tends to
zero at the collapse time (but a Dirac peak is always formed in
the canonical ensemble after $t_{coll}$ as discussed in Appendix
\ref{sec_cold}). We have also evidenced another characteristic
dimension $D=10$ at which the nature of the problem changes. For
$D\ge 10$ the classical spiral characterizing isothermal spheres
in the physical three-dimensional space disappears. However,
since the points in the spiral correspond to unstable states,
that are therefore unphysical, this transition at $D=10$ is not
so critical and indeed the nature of the self-similar collapse
does not show any transition at that dimension. It is interesting
to note that the dependance of the phase diagram in the $(E,T)$
and $(u,v)$ planes with the dimension of space $D$ shows some
resemblance with the dependance of the phase diagram of confined
polytropic spheres with the index $n$ of the polytrope
\cite{poly}. An extension of our study would be to relax the high
friction limit and consider solutions of the Kramers-Poisson
system and other relaxation equations described in Ref.
\cite{csr}. These equations are expected to display qualitatively
similar behaviors than those described here (i.e. gravitational
collapse, finite time singularity, self-similar solutions,...)
but their study appears to be of considerable difficulty since we
now need to consider the evolution of the system in phase space.
We hope to come to that problem in future publications.

\newpage
\appendix

\section{Absence of global maximum of free energy in $D=2$ for $T<T_{c}=GM/4$}
\label{sec_absence}

We give a proof similar to the one given in Ref.~\cite{chavcano} for
$D=3$. In two dimensions, we consider a homogeneous disk of mass $M$
and radius $a$ at temperature $T$. It is easy to show that the total energy
(\ref{maxent8}) of this disk is
\begin{equation}
E=MT+{1\over 2}GM^{2}\Bigl (\ln a-{1\over 4}\Bigr ),
\label{abs1}
\end{equation}
with the convention $\Phi\sim GM\ln r$ at large distances.
According to Eqs.~(\ref{maxent6}) and (\ref{maxent9}), its free
energy reads
\begin{equation}
J=M\ln T-M\ln\biggl ({M\over \pi a^{2}}\biggr )-M-{GM^{2}\over
2T}\Bigl(\ln a-{1\over 4}\Bigr ).
\label{abs2}
\end{equation}
For $a\rightarrow 0$, the free energy behaves like
\begin{equation}
J\sim 2M\biggl (1-{GM\over 4T}\biggr )\ln a.
\label{abs3}
\end{equation}
Therefore, if $T<T_{c}=GM/4$ the free energy goes to $+\infty$ when we
contract the system to a point. This is sufficient to prove the
absence of a global maximum of free energy below $T_{c}$. This also
shows the natural tendency (in a thermodynamical sense) of a
canonical self-gravitating  system to collapse to a Dirac peak for $T<T_{c}$.

\section{The case of cold systems ($T=0$)}
\label{sec_cold}

For $T=0$, Eq.~(\ref{seq}) reduces to
\begin{equation}
\frac{\partial s}{\partial t}=
\left(r\frac{\partial s}{\partial r}+ds\right)s.
\label{scalingt00}
\end{equation}
Looking for a self-similar solution of the form (\ref{dim6s}) and
imposing the conditions (\ref{dim9}) and (\ref{dim11}), we find that
the scaling profile satisfies
\begin{equation}
xS'+\alpha S=\left(xS'+ D S\right)S.
\label{scalingt0}
\end{equation}
Of course, for $T=0$, the exponent $\alpha$ cannot be determined on
dimensional grounds, as the definition $r_0=\sqrt{T/\rho_0}$ is not
relevant anymore. As we will see, $\alpha$ will be determined by
imposing that the scaling solution is analytic. Eq.~(\ref{scalingt0})
can be readily solved leading to the following implicit equation for
$S$:
\begin{equation}
\left(\frac{\alpha}{D}-S(x)\right)^{1-\frac{\alpha}{D}}=K x^{\alpha}S(x),
\label{st0}
\end{equation}
where $K$ is an integration constant. Now, from the definition of $S$,
we expect a small $x$ expansion of the form
$S(x)=S(0)+\frac{1}{2}S''(0)x^2+O(x^4)$, which first implies that
\begin{equation}
S(0)=\frac{\alpha}{D},
\label{S00}
\end{equation}
and that $(x^2)^{1-\alpha/D}\sim x^\alpha$, which finally leads to
\begin{equation}
\alpha=\frac{2D}{D+2} \qquad {\rm and}\qquad K={D+2\over 2}\biggl
({1\over 2}|S''(0)|\biggr )^{D\over D+2}.
\end{equation}
In terms of the scaling function $g(x)$ associated to the mass
profile, Eq.~(\ref{st0}) can be rewritten
\begin{equation}
g(x)=\frac{2x^D}{D+2}-\frac{|S''(0)|}{2}
\left[\frac{D+2}{2}g(x)\right]^{\frac{D+2}{D}},
\label{st0p}
\end{equation}
where $S''(0)<0$ is arbitrary. This leads to the exact large $x$
asymptotic behavior
\begin{equation}
g(x)\sim
\frac{2}{D+2}\left(\frac{4}{(D+2)|S''(0)|}\right)^\frac{D}{D+2}
x^{\frac{D^2}{D+2}}.
\label{sasymt0}
\end{equation}
Moreover, using Eq.~(\ref{dim11}) and (\ref{S00}), we get the exact
universal asymptotic behavior of the central density
\begin{equation}
\rho(0,t)\sim S_D^{-1}(t_{coll}-t)^{-1},
\label{rho0asymt0}
\end{equation}
Finally, we note that the implicit equation (\ref{st0p}) can be written
as a parametric set of equations
\begin{equation}
g(y)=\frac{2}{D+2}y,\qquad x(y)=\left[y+\frac{D+2}{4}|S''(0)|
y^{\frac{D+2}{D}}\right]^{\frac{1}{D}}.
\end{equation}

These results can be obtained by a different, more physical,
method. We have indicated in Ref.~\cite{charosi} that, for $T=0$, the
particles have a deterministic motion with equation
\begin{equation}
{d{\bf r}\over dt}={\bf u}=-\nabla\Phi.
\label{penston1}
\end{equation}
For a spherically symmetrical system, this can be rewritten
\begin{equation}
{dr\over dt}=-{M(r,t)\over r^{D-1}},
\label{penston2}
\end{equation}
where $M(r,t)$ is the mass within $r$. If $a$ denotes the initial
position of the particle located at $r$ at time $t$, we have
\begin{equation}
M(r,t)=M(a,0),
\label{penston2b}
\end{equation}
so Eq.~(\ref{penston2}) can be integrated explicitly in
\begin{equation}
r^{D}=a^{D}-DM(a,0)t.
\label{penston3}
\end{equation}
If $M(a,0)$ behaves like
\begin{equation}
M(a,0)=A(a^{D}-Ba^{D+2})+...,
\label{penston4}
\end{equation}
close to the origin (which is a regular expansion), then
\begin{equation}
M(r,t)=Aa^{D}(1-Ba^{2}), \qquad {\rm with}\qquad
r^{D}=(1-DAt)a^{D}+DABa^{D+2}t.
\label{penston5}
\end{equation}
Introducing the collapse time $t_{coll}=1/DA$ and considering the
limit $t\rightarrow t_{coll}$, we obtain
\begin{equation}
M(r,t)={a^{D}\over Dt_{coll}}, \qquad {\rm with}\qquad r^{D}={1\over
t_{coll}} (t_{coll}-t)a^{D}+Ba^{D+2}.
\label{penston6}
\end{equation}
Introducing the scaling variables
\begin{equation}
x={r\over (t_{coll}-t)^{D+2\over 2D}},\qquad y={1\over
t_{coll}}\biggl\lbrack {a\over (t_{coll}-t)^{1/2}}\biggr \rbrack^{D},
\label{penston7}
\end{equation}
we can put the solution in a parametric form
\begin{equation}
M(r,t)={1\over D}(t_{coll}-t)^{D/2}y, \qquad {\rm with}\qquad
x=(y+Cy^{D+2\over D})^{1\over D},
\label{penston8}
\end{equation}
where $C$ is a constant. At the collapse time $t=t_{coll}$,
\begin{equation}
M(r,t=t_{coll})={1\over DC^{D\over D+2}}r^{D^{2}\over D+2},\qquad
\rho(r,t=t_{coll})={D\over (D+2)S_{D}C^{D\over D+2}}r^{-{2D\over
D+2}}.
\label{penston9}
\end{equation}
These results are of course equivalent to those obtained previously.

We can now use this method to study the evolution of the system for
$t>t_{coll}$ (post-collapse solution). For $t=t_{coll}+\delta t$,
according to Eqs.~(\ref{penston3}) and (\ref{penston9}), the mass
contained inside the sphere of radius $a_{coll}=C^{-1/2}\delta
t^{\frac{D+2}{2D}}$ at $t=t_{coll}$ has collapsed at $r=0$, resulting
in a Dirac peak of weight
\begin{equation}
M(0,t)={1\over DC^{D/2}}(t-t_{coll})^{D/2}.
\label{penston13}
\end{equation}
Note that in a bounded domain the final collapse to a central Dirac peak
containing the whole mass occurs in a finite time after $t_{coll}$.
For $r>0$ (associated to $a>a_{coll}$), one has
\begin{equation}
M(r,t)=M(0,t)+{1\over DC^{D\over D+2}}\left(a^{D^{2}\over D+2}-
a_{coll}^{D^{2}\over D+2}\right),
\label{penston10}
\end{equation}
\begin{equation}
r^{D}=a^{D}\left(1-\left(\frac{a_{coll}}{a}\right)^{\frac{2D}{D+2}}
\right).
\label{penston11}
\end{equation}
Introducing the scaling variables
\begin{equation}
x={r\over a_{coll}},\qquad y=\left(\frac{a}{a_{coll}}\right)
^{\frac{D^2}{D+2}}-1,
\label{penston11b}
\end{equation}
we obtain
\begin{equation}
M(r,t)=M(0,t)(1+y), \qquad {\rm with}\qquad
x=(1+y)^\frac{D^2}{D+2}\left(1-(1+y)^{-{2\over D}}\right)^{1\over D}.
\label{penston12}
\end{equation}
Subtracting the Dirac peak at $r=0$, and considering $x\ll 1$, for
which $y\sim \frac{D}{2}x^D$, we find that the leading contribution to
the mass profile for small $r$ is
\begin{equation}
M(r,t)_{>}\approx \frac{r^D}{2\delta t}.
\label{req0}
\end{equation}
Hence the density profile does not diverge at $r=0^+$ for
$t>t_{coll}$. Instead, the density approaches the constant value
\begin{equation}
\rho(0+,t)= \frac{D}{2S_D\delta t},
\label{rhopluseq0}
\end{equation}
which decreases with time.  The density profile is thus depleted on a
scale $r\sim a_{coll}\sim \delta t^{\frac{D+2}{2D}}$, which increases
with time. For $r\gg a_{coll}$, the density profile remains
essentially unaffected.

In principle, the same phenomenon arises for any $0<T<T_c$: the
density profile obtained at $t_{coll}$ ultimately collapses into a
central Dirac peak at a time $t_{*}>t_{coll}$. This solves the
apparent paradox that the solution at $t=t_{coll}$ has a vanishing
central mass and a finite free energy \cite{charosi}. In fact, if we
allow singular profiles to develop, the evolution continues for
$t>t_{coll}$ and the Dirac peak with infinite free energy (predicted
by statistical mechanics \cite{kiessling}) is formed during the post
collapse regime of our Brownian model \footnote{As discussed in
Sec. \ref{sec_maxent}, the results should be different in the
microcanonical ensemble. We shall reserve for a future communication
the full description of the post-collapse regime.}.  In practice,
degeneracy effects (of quantum or dynamical origin) lead to a finite
small core of finite density, controlled by the maximum allowed
degeneracy \cite{fermions}.

\end{document}